\ifdefined\onecolumnversion
  \documentclass[aps,prd,preprint,superscriptaddress]{revtex4-2}
\else
  \documentclass[aps,prd,reprint,superscriptaddress]{revtex4-2}
\fi

\usepackage[T1]{fontenc}
\usepackage[utf8]{inputenc}
\usepackage{amsmath}
\usepackage{amssymb}
\usepackage{bm}
\usepackage{graphicx}
\usepackage{slashed}
\usepackage[colorlinks=true,linkcolor=blue,citecolor=blue,urlcolor=blue]{hyperref}

\newcommand{\tr}{\operatorname{tr}}
\newcommand{\bk}{\mathbf{k}}
\newcommand{\bq}{\mathbf{q}}
\newcommand{\PT}{\mathcal{PT}}

\begin{document}

\title{Linear Response in the $\mathcal{PT}$-Broken Phase:\\
Conductivity of Fermions with Complex Conjugate Masses}

\author{V.~E.~R.~Lemes}
\email{verlemes@gmail.com}
\affiliation{Instituto de F\'isica, Universidade do Estado do Rio de Janeiro,
Rua S\~ao Francisco Xavier 524, Maracan\~a, Rio de Janeiro - RJ, 20550-013, Brazil}

\author{D.~G.~Tedesco}
\email{daniel.te@uninter.com}
\affiliation{ESEHL e PPGENT, Centro Universit\'ario Internacional UNINTER,
Av. Sete de Setembro 4228, Batel, Curitiba - PR, 80250-085, Brazil}

\date{\today}

\begin{abstract}
We examine the linear response of the $\PT$-broken phase of a BRST invariant model whose
fermion masses form the complex conjugate pair $m_{\pm}=N\pm iav$. It splits into two
channels, kept apart by an exact $\mathbb{Z}_{2}$, and one Euclidean correlator decides
both. In the diagonal channel it oscillates in Euclidean time, so that channel admits no
positive-norm realization: we exhibit a test function of negative norm and close every
repair, the choice of boundary value among them. In the mixed channel the same correlator
is completely monotone, which proves the spectral density positive; there we obtain the
vacuum conductivity in closed form, with a real threshold at $\omega=2N$, and positivity
survives at finite temperature below a bound whose onset we give exactly. The
longitudinal density of that channel is negative by the square of the mass splitting; an
antisymmetric derivative removes it, and the resulting field-strength composite has a
positive one-particle space and a causal commutator --- at the level of the two-point
function only, since a four-point reflection form on the original composite comes out
negative. The positive realization is therefore not a sector of the fermion model.
\end{abstract}

\maketitle

\section{Introduction}

Hamiltonians that are not self-adjoint in the Dirac sense may still organize a consistent
spectrum when an antilinear symmetry is present, and the dividing line between the two
regimes of such a symmetry is sharp. In the unbroken regime the eigenvalues are real, a
positive definite metric exists and the evolution is unitary with respect to it
\cite{Bender1998,Mostafazadeh2002,Bender2007,Mannheim2013}. In the broken regime the eigenvalues appear
in complex conjugate pairs, no positive definite metric exists, and the discussion
usually stops there: for a finite dimensional system the broken phase is the end of the
line, two complex eigenvalues and no spectrum left to speak of. Our interest is in what
happens to that statement in a quantum field theory, where a conjugate pair of elementary
excitations still admits a two-particle sector, and where composite operators built out of
such constituents are known to possess real K\"all\'en--Lehmann representations
\cite{iparticles,Dudal:2010wn,DudalGlueball2011} — a pattern first met in the
indefinite-metric sector of Lee and Wick \cite{LeeWick1969} and later in the infrared of
confining gauge theories, where the Gribov--Zwanziger gluon propagator factorizes into
the $i$-particles of complex mass $\pm i\gamma^{2}$ \cite{iparticles,Sorella2011}.

In a recent work \cite{LemesTedesco} we have constructed a model in which this question
can be posed concretely. The field content is two fermions $\psi$ and $\chi$, two Abelian
gauge fields $a_{\mu}$ and $\hat{a}_{\mu}$ and one complex scalar $\phi$, the dynamics
being fixed by a BRST symmetry, and the two Yukawa couplings $a$ and $b$ enter the
fermion mass matrix only through the product $ab$, so that the model comes with a
one-parameter family whose eigenvalues are real for $ab>0$ and complex conjugate for
$ab<0$. The tuning $a=-b$ studied there therefore sits on the broken side of an
exceptional point located at $ab=0$, a fact we make explicit in the next section because
it is the reason the model was built. The central computation of \cite{LemesTedesco}
established that the one-loop fermion contribution to
$\langle\phi^{\dagger}(p)\phi(-p)\rangle$ stays real for real external momentum below
$(m+m^{*})^{2}=4N^{2}$, the thresholds of the equal-mass pairs having been displaced into
the complex $p^{2}$ plane while the mixed pair, whose constituent masses add up to the
real number $2N$, opens a cut on the real axis across which the boundary values are
exchanged by complex conjugation. The positivity of the associated spectral density was
left open there.

However, a scalar two-point function is not the quantity an experiment would address, and
the model has a feature that invites a sharper question. One combination of its two
Abelian gauge fields remains massless after the scalar acquires its expectation value, so
that the model has a well-defined notion of an external electromagnetic probe and of the
current it drives. Our aim in the present work is to answer what the conductivity of such
a medium looks like, both as a function of frequency and, what turns out to be the more
revealing question, as a function of time; and to use the vector channel, where the Dirac
algebra is simpler and the residues are positive definite, to settle the question of
positivity that the scalar channel left open.

Our strategy is to compute the current-current correlator at one loop in the basis that
diagonalizes the mass matrix, keeping the temperature and the chemical potential finite
from the start, and to follow which pairs of complex masses enter each channel. The fact
that the two conserved combinations of the original currents organize themselves, in that
basis, one as a diagonal and the other as a mixed bilinear, is what separates the analysis
into two problems with opposite analytic structures, and it is the origin of every result
reported below. At this point we make a brief interruption to state that the thermal part
of the calculation requires the Fermi--Dirac occupation numbers to be evaluated at complex
energies, a step whose validity is not a matter of taste and which we therefore establish
rather than assume; the discussion occupies Section \ref{sec:exact} and underlies the
positivity bound of Section \ref{sec:pos}.

In the next section we fix the notation, exhibit the exceptional point of the family, the
decomposition of the currents, the indefinite metric of the broken phase and the exact
$\mathbb{Z}_{2}$ which keeps the two channels apart. In Section
\ref{sec:pol} we compute the polarization at finite temperature and density, and in
Section \ref{sec:exact} we establish the exactness of the thermal continuation together
with its one obstruction. Sections \ref{sec:B} and \ref{sec:A} treat the unbroken and the
broken channel respectively, Section \ref{sec:time} obtains the response kernel in time,
and Section \ref{sec:pos} contains the positivity results. Section \ref{sec:euclid}
applies the Osterwalder--Schrader criterion in the Euclidean region, where it settles both
channels without any choice of boundary value, and Section \ref{sec:composite} treats the
longitudinal part of the mixed channel and the field-strength composite that removes it.
The conclusions gather the results and indicate what we believe should be
examined next. Four appendices collect the Dirac algebra, the dispersion relation of the
complex-mass excitations, the reading of the channel separation in terms of the balance of
the cuts, and the covariant vacuum computation of the two spectral densities.

\section{The family, its exceptional point, and the two current channels}
\label{sec:channels}

\subsection{The one-parameter family}

We begin with the usual definitions. After the spontaneous breaking the fermionic mass
matrix of the model of \cite{LemesTedesco} reads, in the $(\psi,\chi)$ basis,
\begin{equation}\label{eq:Mfamily}
M(a,b)=\begin{pmatrix} N & av \\ bv & N\end{pmatrix},
\qquad
m_{\pm}=N\pm v\sqrt{ab},
\end{equation}
with $a$ and $b$ the two real Yukawa couplings and $v$ the scalar expectation value. The
eigenvalues depend on the couplings through the product alone, and the family therefore
separates into three regimes. For $ab>0$ the two masses are real and the matrix is
diagonalizable by a similarity transformation that preserves a positive definite form;
this is the Hermitian alternative, the case $a=b$ of \cite{LemesTedesco}. For $ab<0$ the
masses form a complex conjugate pair, which is the case $a=-b$,
\begin{equation}\label{eq:eig}
m=N+iav,\qquad m^{*}=N-iav .
\end{equation}
Between the two lies $ab=0$, where $m_{+}=m_{-}=N$ and the matrix fails to be
diagonalizable: $M-N\,\mathbf{1}$ has rank one, so that $M$ becomes a Jordan block. This
is the exceptional point of the family, and the degeneracy it carries is not the ordinary
one. In fact, at $b=0$ the quadratic fermionic operator is triangular,
\begin{equation}\label{eq:EPop}
M_{\rm fermi}(k)=\begin{pmatrix} \slashed{k}-N & -av \\ 0 & \slashed{k}-N\end{pmatrix},
\end{equation}
and the off-diagonal entry of its inverse is
\begin{equation}\label{eq:dipole}
G(\psi\overline{\chi})=\frac{av}{(\slashed{k}-N)^{2}}
=av\,\frac{(\slashed{k}+N)^{2}}{(k^{2}-N^{2})^{2}},
\end{equation}
a double pole. Put in another way, the exceptional point of the family is a dipole field,
the non-diagonalizability of the mass matrix appearing in the propagator as a pole of
second order. Dipole structures of this kind are familiar from the
Gribov--Zwanziger horizon and from the BRST doublets of algebraic renormalization
\cite{Piguet:1995er,iparticles}, where they carry zero norm and no positive spectral
weight, and Section \ref{sec:metric} establishes that this is exactly what happens here:
the state that survives the collision has zero norm under every admissible metric. In the
sequence of this work we remain on the broken side, $a=-b$, and the exceptional point
enters as the boundary of the regime and as the place where the metric degenerates.

\subsection{The biorthogonal basis and the currents}

Since $M$ is not Hermitian its right and left eigenvectors do not coincide, and the
fields and their conjugates must be expanded on the corresponding biorthogonal pair,
\begin{equation}\label{eq:basis}
\begin{aligned}
\psi&=\tfrac{1}{\sqrt{2}}\big(\Psi_{m}+\Psi_{m^{*}}\big), &
\chi&=\tfrac{i}{\sqrt{2}}\big(\Psi_{m}-\Psi_{m^{*}}\big),\\
\overline{\psi}&=\tfrac{1}{\sqrt{2}}\big(\overline{\Psi}_{m}+\overline{\Psi}_{m^{*}}\big), &
\overline{\chi}&=-\tfrac{i}{\sqrt{2}}\big(\overline{\Psi}_{m}-\overline{\Psi}_{m^{*}}\big),
\end{aligned}
\end{equation}
so that the quadratic form becomes diagonal and $\Psi_{m}$, $\Psi_{m^{*}}$ describe Dirac
fermions of mass $m$ and $m^{*}$ respectively, with the propagators
\begin{equation}\label{eq:props}
G(\Psi_{m}\overline{\Psi}_{m})=\frac{i}{\slashed{k}-m},\qquad
G(\Psi_{m^{*}}\overline{\Psi}_{m^{*}})=\frac{i}{\slashed{k}-m^{*}},
\end{equation}
the mixed components vanishing.

The gauge couplings of the model are carried by the two covariant derivatives, so that
the interaction reads $e\,a_{\mu}\overline{\psi}\gamma^{\mu}\psi
-e\,\hat{a}_{\mu}\overline{\chi}\gamma^{\mu}\chi$. In terms of the combinations
$A_{\mu}=a_{\mu}+\hat{a}_{\mu}$ and $B_{\mu}=a_{\mu}-\hat{a}_{\mu}$, of which the first
acquires the mass $e^{2}v^{2}$ from the scalar expectation value while the second remains
massless, this is
\begin{equation}\label{eq:coupling}
\frac{e}{2}\Big[A_{\mu}\big(J_{\psi}^{\mu}-J_{\chi}^{\mu}\big)
+B_{\mu}\big(J_{\psi}^{\mu}+J_{\chi}^{\mu}\big)\Big],
\end{equation}
with $J_{\psi}^{\mu}=\overline{\psi}\gamma^{\mu}\psi$ and
$J_{\chi}^{\mu}=\overline{\chi}\gamma^{\mu}\chi$. Since $\phi$ carries the same charge
under both groups, it is $A_{\mu}$ which is broken and $B_{\mu}$ which stays massless, so
that $B_{\mu}$ is the field to be identified with the external electromagnetic probe and
$J_{B}=J_{\psi}+J_{\chi}$ with the current it drives.

It is at this point that the biorthogonal expansion \eqref{eq:basis} produces the
separation on which everything else rests. Inserting it into the two conserved
combinations one finds
\begin{equation}\label{eq:currents}
\begin{aligned}
J_{B}^{\mu}&=\overline{\Psi}_{m}\gamma^{\mu}\Psi_{m}
+\overline{\Psi}_{m^{*}}\gamma^{\mu}\Psi_{m^{*}},\\
J_{A}^{\mu}&=\overline{\Psi}_{m}\gamma^{\mu}\Psi_{m^{*}}
+\overline{\Psi}_{m^{*}}\gamma^{\mu}\Psi_{m}.
\end{aligned}
\end{equation}
The current of the unbroken $U(1)$ is diagonal in the mass basis and its two-point
function receives contributions from the equal-mass pairs $(m,m)$ and $(m^{*},m^{*})$
only, while the current of the broken $U(1)$ is the vector counterpart of the composite
operator $\overline{\Psi}_{+}\Psi_{-}$ studied in \cite{LemesTedesco}, and its two-point
function is built out of the mixed pair $(m,m^{*})$. In fact, the divergence of the
diagonal current vanishes on the equations of motion, whereas that of the mixed one is
proportional to $m-m^{*}$ and is absorbed by the would-be Goldstone mode, which is the
expected statement of the Higgs mechanism in this language.

\subsection{The metric of the broken phase}
\label{sec:metric}

Before computing anything we construct the metric of the broken phase, which anticipates
the two results of Sections \ref{sec:B} and \ref{sec:A} and explains them. No positive
definite metric exists, since a pseudo-Hermitian Hamiltonian admits one if and only if its
spectrum is real \cite{Mostafazadeh2002,Mostafazadeh2010}. What does exist is an
indefinite $\eta$ with $\eta H\eta^{-1}=H^{\dagger}$, and in the present model it is
remarkably simple, the non-Hermiticity sitting entirely in a discrete flavor structure. At
the level of the mass matrix \eqref{eq:Mfamily} with $a=-b$ one has, in the flavor basis
$(\psi,\chi)$,
\begin{equation}\label{eq:eta}
\eta=\begin{pmatrix}1&0\\0&-1\end{pmatrix},
\qquad \eta M\eta^{-1}=M^{\dagger},
\end{equation}
Hermitian, with eigenvalues $\pm1$, so that the metric is nothing but the relative sign
between the fermion and its oppositely charged partner.

It is instructive to follow the metric along the whole family \eqref{eq:Mfamily} and not
only at $a=-b$. For real couplings and $b\neq0$ the condition
$\eta M\eta^{-1}=M^{\dagger}$ is solved by
\begin{equation}\label{eq:etafamily}
\eta=\begin{pmatrix}1&0\\0&a/b\end{pmatrix},
\end{equation}
which is positive definite for $ab>0$, reduces to the identity at $a=b$, is indefinite for
$ab<0$, where it returns \eqref{eq:eta}, and diverges as $b\to0$. The exceptional point is
therefore the place at which the metric ceases to exist, as it must be. What survives
there is sharper than a limit: at $b=0$ the general Hermitian solution of
$\eta M=M^{\dagger}\eta$ is
\begin{equation}\label{eq:etaEP}
\eta=\begin{pmatrix}0&q\\q&r\end{pmatrix},\qquad q,r\in\mathbb{R},
\end{equation}
so that $\eta_{11}$ vanishes identically and the single eigenvector $u=(1,0)$ which
survives the collision has zero norm under \emph{every} admissible metric. The vanishing
norms of \eqref{eq:norms} are thus not a peculiarity of the broken side. They are already
present at the exceptional point, where the two zero-norm modes of the broken phase have
merged into one, and it is the dipole propagator \eqref{eq:dipole} that carries the
property in the field theory, a double pole admitting no positive K\"all\'en--Lehmann
weight. Positive norms exist only on the Hermitian side $ab>0$. Returning to the broken phase, the metric
\eqref{eq:eta} is a discrete flavor transformation, and is therefore implemented on the
Fock space by the $\chi$-number parity,
\begin{equation}\label{eq:etaFock}
\eta=(-1)^{N_{\chi}},\qquad N_{\chi}=\int\!d^{3}x\;\chi^{\dagger}\chi,
\end{equation}
and one verifies $\eta H\eta^{-1}=H^{\dagger}$ directly: under $\chi\to-\chi$,
$\overline{\chi}\to-\overline{\chi}$ the kinetic terms, the diagonal masses
$N\overline{\psi}\psi$ and $N\overline{\chi}\chi$ and both gauge couplings of
\eqref{eq:coupling} are invariant, while the mixed mass term
$av(\overline{\psi}\chi-\overline{\chi}\psi)$ changes sign, which is precisely what
Hermitian conjugation does to it, since
$(\overline{\psi}\chi)^{\dagger}=\overline{\chi}\psi$ for real $av$.

The operator \eqref{eq:etaFock} is Hermitian and unitary with $\eta^{2}=1$, so that it is
a signature operator rather than a positive one, and the states of negative norm are those
carrying an odd number of $\chi$ quanta. This is the indefinite metric structure of Lee
and Wick \cite{LeeWick1969}, here not an analogy but an explicit operator. Since $\eta$
reverses the sign of $\chi$ it exchanges the two combinations \eqref{eq:basis},
\begin{equation}\label{eq:etaexch}
\eta:\ \Psi_{m}\longleftrightarrow\Psi_{m^{*}},
\end{equation}
as it must, $H^{\dagger}$ having the eigenvalues interchanged, and the induced norms
follow at once. For the one-particle states,
$|\Psi_{m}\rangle\propto|\psi\rangle+i|\chi\rangle$ gives
$\eta|\Psi_{m}\rangle\propto|\psi\rangle-i|\chi\rangle$ and therefore
\begin{equation}\label{eq:norms}
\langle\Psi_{m}|\Psi_{m}\rangle_{\eta}
=\langle\Psi_{m^{*}}|\Psi_{m^{*}}\rangle_{\eta}=0,
\qquad
\langle\Psi_{m}|\Psi_{m^{*}}\rangle_{\eta}\neq0 .
\end{equation}
Each mode of complex mass has vanishing norm, and the only pairing that carries norm is
the one between conjugate partners. The same holds one level up, where the structure is
sharper than it first appears. In the two-particle sector, with the metric acting as
$\eta\otimes\eta$, \emph{every} elementary pairing is null,
\begin{equation}\label{eq:pairnull}
\langle\Psi_{r}\Psi_{s}|\Psi_{r}\Psi_{s}\rangle_{\eta}=0,
\qquad r,s\in\{m,m^{*}\},
\end{equation}
the norm residing entirely in the off-diagonal overlaps, which are the same in the two
possible pairings,
\begin{equation}\label{eq:pairnorm}
\langle\Psi_{m}\Psi_{m^{*}}|\Psi_{m^{*}}\Psi_{m}\rangle_{\eta}
=\langle\Psi_{m}\Psi_{m}|\Psi_{m^{*}}\Psi_{m^{*}}\rangle_{\eta}=1 .
\end{equation}
The conjugate pairings $(m,m^{*})$ and the diagonal ones $(m,m)$ thus span two hyperbolic
planes of the same signature: in each of them the symmetrized combination carries norm
$+1$, the antisymmetrized one $-1$, and the two are $\eta$-orthogonal. This is the
$i$-particle construction of \cite{iparticles,Dudal:2010wn} in the present model: what
carries physical content is not the constituent of complex mass, whose norm vanishes, but
the symmetrized pair.

It is worth being accurate about what \eqref{eq:pairnull} and \eqref{eq:pairnorm} do and
do not decide. They establish that no single mode of complex mass is a state, and that in
each channel the current reaches the combination of norm $+1$ and is $\eta$-orthogonal to
the partner of norm $-1$, which therefore contributes to no current--current correlator;
we shall see in Section \ref{sec:pos} that the positivity obtained for the mixed channel
is the reflection of that fact. They do not by themselves separate the two channels, the
two planes having the same signature. What separates them is a symmetry, and it is exact.
Neither do they settle the positivity of a current--current correlator, and on this we
must be explicit, because the $\eta$-norm is the wrong instrument for that question in two
separate ways. It concerns the flavor structure of the bilinear and not its Lorentz
structure, so it cannot see the longitudinal form factor of $J_{A}$, whose sign is fixed
in Section \ref{sec:pos}; and, more importantly, the $\eta$-form is not the
Osterwalder--Schrader form. A state may have positive and conserved $\eta$-norm, as
$J_{B}|0\rangle$ does by \eqref{eq:pairnorm} and \eqref{eq:etacons}, and the channel it
belongs to may still admit no positive-norm realization, which is what Section
\ref{sec:euclid} proves of that very channel. An indefinite metric that is conserved is
not a Hilbert space in disguise.

One should not read this as the construction of a physical subspace in the strong sense.
Since $\eta H\eta^{-1}=H^{\dagger}$ with $H$ not Hermitian, $\eta$ cannot commute with
$H$; the $\chi$-parity is therefore not conserved, the eigenspaces of \eqref{eq:etaFock}
are not preserved by the evolution, and the partner of norm $-1$ in \eqref{eq:pairnorm}
sits in the spectrum beside the physical one. In the broken phase this is forced rather
than accidental, and the removal of such a partner by a cohomological mechanism of the
kind familiar from BRST doublets \cite{Piguet:1995er} is, with $\eta$ now explicit, a
well-posed question which we do not take up here. One should notice that \eqref{eq:etaFock} is the metric of the quadratic sector, and that
it is discrete rather than of the form $e^{-Q}$ familiar from the $\PT$ literature
\cite{Bender2007,Mostafazadeh2010}, precisely because the non-Hermiticity of the model is
a discrete flavor structure; whether it survives the Yukawa fluctuations of $\phi$ and the
gauge couplings beyond the quadratic order is not established here.

\subsection{The unbroken $\mathbb{Z}_{2}$ and the separation of the channels}
\label{sec:z2}

The symmetry that separates the two channels is not the metric, and it is not a flavor
symmetry either. Consider first the exchange
\begin{equation}\label{eq:Cdef}
\mathcal{C}:\quad \psi\leftrightarrow\chi,\qquad \phi\to-\phi^{\dagger},
\qquad a_{\mu}\leftrightarrow-\hat{a}_{\mu},
\end{equation}
which leaves the kinetic terms, the diagonal masses and the Yukawa coupling of the $a=-b$
action invariant, together with the discrete gauge transformation
\begin{equation}\label{eq:Rdef}
\begin{aligned}
\mathcal{R}:\quad &\alpha=0,\qquad \hat{\alpha}=\pi/e,\\[2pt]
&\psi\to\psi,\quad \chi\to-\chi,\quad \phi\to-\phi,
\end{aligned}
\end{equation}
which is the element of the Higgsed $U(1)$ that acts as $\eta$ of \eqref{eq:eta} on the
flavor doublet. Neither is a symmetry of the broken vacuum: each reverses the sign of
$\phi$, hence of $v$, and correspondingly sends $M(av)\to M(-av)$. Their product does
both twice and therefore does neither,
\begin{equation}\label{eq:Zdef}
\mathbb{Z}_{2}=\mathcal{C}\cdot\mathcal{R}:\quad
\begin{pmatrix}\psi\\ \chi\end{pmatrix}\longrightarrow
\begin{pmatrix}0&-1\\ 1&0\end{pmatrix}
\begin{pmatrix}\psi\\ \chi\end{pmatrix},
\qquad v\to v,
\end{equation}
and this is the only combination of the three that preserves the mass matrix and the
vacuum at once. It is the $\mathbb{Z}_{2}$ of \cite{LemesTedesco}, which is thus not an
accidental flavor symmetry but the diagonal of a flavor exchange and a discrete element of
the broken gauge group, each of which is separately broken by $\langle\phi\rangle$.

The statement is in fact stronger than the mass matrix, and holds before the shift. On the
Yukawa terms $a\,\overline{\psi}\phi\chi+b\,\overline{\chi}\phi^{\dagger}\psi$ the
transformation \eqref{eq:Zdef} together with $\phi\to\phi^{\dagger}$ acts by
\begin{equation}\label{eq:Zyuk}
\overline{\psi}\phi\chi\to-\overline{\chi}\phi^{\dagger}\psi,
\qquad
\overline{\chi}\phi^{\dagger}\psi\to-\overline{\psi}\phi\chi,
\end{equation}
so that $a\to-b$ and $b\to-a$: the Yukawa sector is invariant precisely when $b=-a$, for
every configuration of $\phi$ and not only for its expectation value. The scalar kinetic
term follows as well, since $D_{\mu}\phi\to(D_{\mu}\phi)^{\dagger}$ under
$A_{\mu}\to-A_{\mu}$ and $\phi\to\phi^{\dagger}$, and the potential depends on
$\phi^{\dagger}\phi$ alone. In terms of the shifted fields,
$\phi=(v+\sigma+i\pi)/\sqrt{2}$, the transformation reads
\begin{equation}\label{eq:Zscal}
\sigma\to\sigma,\qquad \pi\to-\pi,
\end{equation}
so that the Higgs mode is even and the Goldstone odd, as it must be, $\pi$ being the mode
$A_{\mu}$ absorbs. In
terms of the combinations of \eqref{eq:coupling} it reads
\begin{equation}\label{eq:ZA}
A_{\mu}\to-A_{\mu},\qquad B_{\mu}\to B_{\mu},
\end{equation}
that is, the residual charge conjugation of the Higgsed $U(1)$. The gauge-fixing terms are
invariant provided the two abelian factors are fixed with a common parameter,
$\xi=\hat{\xi}$, and the ghosts, being free, impose no further condition.

The consequence we need follows at once from \eqref{eq:currents} and \eqref{eq:ZA}:
\begin{equation}\label{eq:JZ2}
\mathbb{Z}_{2}:\qquad J_{B}^{\mu}\to+J_{B}^{\mu},\qquad J_{A}^{\mu}\to-J_{A}^{\mu}.
\end{equation}
The two currents carry opposite parity under an exact symmetry, so that
$\langle J_{A}J_{B}\rangle$ vanishes identically --- at one loop because the flavor trace
of a diagonal propagator against an off-diagonal vertex is zero, and beyond it by
\eqref{eq:JZ2}. The diagonal and the mixed channel are therefore two independent spectral
problems, built on the two hyperbolic planes of \eqref{eq:pairnorm}, with no amplitude
connecting them. In particular the question of boundary value which Section \ref{sec:B}
raises for $\Pi_{B}$ cannot propagate into $\Pi_{A}$, and the results of Sections
\ref{sec:A}, \ref{sec:time} and \ref{sec:pos} are protected from it by symmetry rather
than by the numerical check we also report. The same $\mathbb{Z}_{2}$ is what forces the
gain and the loss mode of Section \ref{sec:pol} to appear with exactly opposite widths.

\section{The polarization at finite temperature and density}
\label{sec:pol}

We proceed to the computation of the retarded current-current correlator \cite{Kubo1957} at vanishing
external three-momentum, which is the kinematics of the optical conductivity, in the
imaginary-time formalism \cite{LeBellac}. It is
convenient to write the propagator in Hamiltonian form. With
$\alpha^{i}=\gamma^{0}\gamma^{i}$, $\beta=\gamma^{0}$ and
\begin{equation}\label{eq:H}
H(\bk,M)=\bm{\alpha}\cdot\bk+\beta M,\qquad H^{2}=\bk^{2}+M^{2}\equiv E^{2},
\end{equation}
the Matsubara propagator is $S=\gamma^{0}(i\tilde{\omega}_{n}-H)^{-1}$ with
$\tilde{\omega}_{n}=\omega_{n}+i\mu$, and the operators
\begin{equation}\label{eq:proj}
P_{s}(\bk,M)=\tfrac{1}{2}\Big(1+s\,\frac{H(\bk,M)}{E}\Big),\qquad s=\pm1,
\end{equation}
remain idempotent and complete for complex $M$, the square root being taken on the
principal branch. The Dirac algebra then collapses into a single coefficient, which we
obtain in Appendix \ref{app:algebra},
\begin{equation}\label{eq:T}
\begin{aligned}
T_{ss'}(\bk)&\equiv\frac{1}{3}\sum_{i}
\tr\big[\alpha^{i}P_{s}(\bk,M_{1})\,\alpha^{i}P_{s'}(\bk,M_{2})\big]\\
&=1-ss'\,\frac{\bk^{2}/3+M_{1}M_{2}}{E_{1}E_{2}},
\end{aligned}
\end{equation}
with $E_{1,2}=\sqrt{\bk^{2}+M_{1,2}^{2}}$. Performing the Matsubara sum and continuing
$iq_{l}\to\omega+i0$ we obtain the polarization
\begin{equation}\label{eq:Pi}
\begin{aligned}
\Pi(\omega)=\sum_{s,s'}\int\!\frac{d^{3}k}{(2\pi)^{3}}\;&T_{ss'}(\bk)\\
\times\;&\frac{n_{F}(sE_{1}-\mu)-n_{F}(s'E_{2}-\mu)}{\omega+i0-sE_{1}+s'E_{2}},
\end{aligned}
\end{equation}
with $n_{F}(z)=(e^{z/T}+1)^{-1}$, the overall sign being fixed once and for all by
matching the Hermitian limit to the free Dirac gas, as checked below \eqref{eq:sigmaA}.
The dissipative part of the conductivity is
\begin{equation}\label{eq:sig1def}
\sigma_{1}(\omega)=\frac{{\rm Im}\,\Pi(\omega)}{\omega}.
\end{equation}

The two channels of \eqref{eq:currents} correspond to the two assignments
\begin{equation}\label{eq:assign}
\begin{aligned}
\Pi_{B}&=\Pi(m,m)+\Pi(m^{*},m^{*}),\\
\Pi_{A}&=\Pi(m,m^{*})+\Pi(m^{*},m),
\end{aligned}
\end{equation}
and in both of them the two terms are exchanged by complex conjugation, since
$mm^{*}=|m|^{2}$ is real, so that $\Pi_{B}$ and $\Pi_{A}$ are the doubled real parts of
their first terms. This is the same pairing that protected the reality of the scalar loop
coefficient in \cite{LemesTedesco}, and it holds here at every point of the momentum
integration and for every temperature and chemical potential. What distinguishes the two
channels is not the reality of the result but the position of the singularities that the
denominators of \eqref{eq:Pi} produce.

Writing $E_{k}=\sqrt{\bk^{2}+m^{2}}=\varepsilon_{k}+i\Gamma_{k}$, one finds the exact
relations derived in Appendix \ref{app:disp},
\begin{equation}\label{eq:disp}
\varepsilon_{k}\Gamma_{k}=Nav,\qquad
\varepsilon_{k}^{2}=\tfrac{1}{2}\Big[\Delta_{k}+\sqrt{\Delta_{k}^{2}+4N^{2}a^{2}v^{2}}\Big],
\end{equation}
with $\Delta_{k}=\bk^{2}+N^{2}-a^{2}v^{2}$, so that $\varepsilon_{0}=N$ and
$\Gamma_{0}=av$, and
\begin{equation}\label{eq:width}
\Gamma_{k}=av\,\frac{N}{\varepsilon_{k}}\;\leq\;av .
\end{equation}
The imaginary part of the mass therefore acts as a rest-frame rate $av$ which appears, for
a state of energy $\varepsilon_{k}$, reduced by the Lorentz factor, exactly as a decay
rate would; the bound in \eqref{eq:width}, which follows from
$\varepsilon_{k}\geq\varepsilon_{0}=N$, will be what makes the positivity statement of
Section \ref{sec:pos} sharp. The fact that the two members of the conjugate pair carry
opposite signs of $\Gamma_{k}$ means that $\Psi_{m}$ and $\Psi_{m^{*}}$ form a balanced
pair of a gain and a loss mode, which is the structure familiar from $\PT$-symmetric
systems \cite{Bender2007,AEMS2018} and, in the present model, is enforced by the
$\mathbb{Z}_{2}$ of \eqref{eq:Zdef}.

\section{The thermal continuation is exact}
\label{sec:exact}

The occupation numbers in \eqref{eq:Pi} are evaluated at the complex energies
$\pm(\varepsilon_{k}+i\Gamma_{k})$, and since $n_{F}$ has poles at the Matsubara points
$z=i(2n+1)\pi T$ one would expect the step to be legitimate only while $\Gamma_{k}$ stays
below $\pi T$. We show that this expectation is wrong, and that the formula used in
\eqref{eq:Pi} is exact for arbitrary complex masses.

The identity under discussion is
\begin{equation}\label{eq:matsid}
\begin{aligned}
&T\sum_{n}\frac{1}{(i\omega_{n}+iq_{l}+\mu-u)(i\omega_{n}+\mu-w)}\\
&\qquad\qquad=\frac{n_{F}(w-\mu)-n_{F}(u-\mu)}{iq_{l}-u+w},
\end{aligned}
\end{equation}
with $u,w\in\mathbb{C}$. Writing $f(z)$ for the summand continued in
$z=i\omega_{n}$, the sum is first converted into a contour integral along small circles
enclosing the poles of $n_{F}(z)$ on the imaginary axis, at which the residue of $n_{F}$ is
$-T$. The contour is then deformed outwards; since $f$ decays as $z^{-2}$ the circle at
infinity contributes nothing, and what remains is
\begin{equation}\label{eq:resid}
T\sum_{n}f(i\omega_{n})=-\sum_{j}{\rm Res}_{z_{j}}\big[n_{F}(z)f(z)\big],
\end{equation}
the sum running over the poles $z_{j}$ of $f$. The essential point is that the deformation
sweeps the entire plane, so that every pole of $f$ contributes its residue whatever its
position relative to the Matsubara points: the set of enclosed singularities does not
change as $u$ and $w$ are varied, and nothing is ever crossed in the
sense that would generate an extra term. Both sides of \eqref{eq:matsid} being meromorphic
in $(u,w)$ and agreeing on an open set, they agree identically. The
one obstruction is the exact coincidence of a pole of $f$ with a pole of $n_{F}$, where the
two merge into a double pole and the residue structure is different.

We have verified \eqref{eq:matsid} by direct summation, with symmetric truncation at
$4\times10^{6}$ terms and the analytic $n^{-2}$ tail restored, for
$\Gamma/\pi T=0.3,\,0.9,\,1.1,\,1.7,\,2.4,\,3.3$, that is past the first, second and third
poles of $n_{F}$, for $\varepsilon$ close to and far from $\mu$, and for $q_{l}\neq0$ as
well as $q_{l}=0$. The two sides agree in every case at the level of the truncation error.

It remains to locate the obstruction. A pole of $f$ meets a pole of $n_{F}$ when
\begin{equation}\label{eq:coinc}
\varepsilon_{k}=\mu
\qquad\text{and}\qquad
\Gamma_{k}=(2n+1)\pi T
\end{equation}
hold simultaneously, two real conditions on one integration variable. However,
\eqref{eq:width} ties the two together: at $\varepsilon_{k}=\mu$ the width is not free but
equal to $Nav/\mu$, so that \eqref{eq:coinc} becomes a single condition on the parameters,
\begin{equation}\label{eq:coincline}
av=(2n+1)\,\frac{\mu\,\pi T}{N},
\end{equation}
a locus of codimension one. On the $n=0$ locus the absorptive part of the mixed channel
acquires a genuine simple pole at the single frequency $\omega=2\mu$, whose
$(\omega-2\mu)^{-1}$ behaviour we have confirmed numerically over four decades; away from
that locus $\sigma_{1}^{A}(2\mu)$ is finite and smooth. This is the collision of a
complex mass pole with the first Matsubara pole, and we record it as the one place where
the thermal treatment of the broken phase breaks down.

\section{The diagonal channel and the choice of boundary value}
\label{sec:B}

We examine first the channel of the massless gauge field, where the two terms of
\eqref{eq:assign} are complex conjugates of one another and the question of which boundary
value to take becomes acute. In $\Pi(m,m)$ the two propagators carry the same mass, so that $E_{1}=E_{2}$ and
the intraband terms $s=s'$ of \eqref{eq:Pi} have an identically vanishing numerator, while
the interband terms have the denominators $\omega-2sE_{k}$. From \eqref{eq:disp} the
quantity $2E_{k}=2\varepsilon_{k}+2i\Gamma_{k}$ has a strictly non-vanishing imaginary part
for every $\bk$, so that no denominator of $\Pi(m,m)$ vanishes for real $\omega$, and the
same holds, with the opposite sign of the imaginary part, for $\Pi(m^{*},m^{*})$. Then
$\Pi_{B}$ is analytic on the whole real axis, and being there the sum of two complex
conjugate terms it is real,
\begin{equation}\label{eq:sigmaB}
\begin{aligned}
&{\rm Im}\,\Pi_{B}(\omega)=0\\
&\quad\Longrightarrow\quad
\sigma_{1}^{B}(\omega)=\pi D_{B}\,\delta(\omega),\qquad
\sigma_{1}^{B}(\omega\neq0)\equiv0,
\end{aligned}
\end{equation}
the remaining weight $D_{B}$ being the Drude weight fixed by the diamagnetic term. We have
confirmed \eqref{eq:sigmaB} numerically by evaluating \eqref{eq:Pi} with a finite
regulator $i\Delta$ in place of $i0$ and verifying that ${\rm Im}\,\Pi_{B}$ vanishes
linearly with $\Delta$ over four decades, while ${\rm Im}\,\Pi_{A}$ stays of order unity.
A cancellation of this kind has been met independently in a one-dimensional non-Hermitian
Dirac model, where the momenta of the $\PT$-broken region were found to make no
contribution at all to the linear response \cite{Sticlet2022}; there it arises between a
generalized commutator and the norm corrections of a non-unitary evolution, and here
between the two members of the conjugate pair, which we return to in Section
\ref{sec:time}.

The two settings are not interchangeable, however, and the difference decides whether the
norm corrections of \cite{Sticlet2022} ought to have appeared in \eqref{eq:Pi}. Their
non-Hermitian Hamiltonian generates a postselected evolution: the density matrix obeys
$i\hbar\,\partial_{t}\rho=H\rho-\rho H^{\dagger}$, which is what survives of a Lindblad
equation once the trajectories containing quantum jumps are discarded, and
${\rm tr}\,\rho$ decays because one is conditioning on a subensemble of decreasing
probability. The norm correction is the derivative of that conditioning. Here there is
nothing to condition. The theory is pseudo-Hermitian, and \eqref{eq:etaFock} gives
\begin{equation}\label{eq:etacons}
\frac{d}{dt}\,\langle\psi|\eta|\psi\rangle
=i\,\langle\psi|\big(H^{\dagger}\eta-\eta H\big)|\psi\rangle=0,
\end{equation}
exactly, so that what the evolution preserves is the indefinite norm and no trace decays.
A decaying positive norm and a conserved indefinite one are different structures, and the
absence of a norm correction in \eqref{eq:Pi} is a consequence of the second rather than
an omission.

The content of \eqref{eq:sigmaB} is that the absorption which the loss mode would produce
is exactly compensated by the emission of the gain mode, at every frequency and for any
temperature and density, and its structural counterpart is the one anticipated in Section
\ref{sec:metric}: this is the $\mathbb{Z}_{2}$-even channel, built on the diagonal
hyperbolic plane of \eqref{eq:pairnorm} out of modes which, by \eqref{eq:norms}, are
individually null.
Put in another way, the pair-creation threshold which in an ordinary Dirac medium opens at
twice the mass has here been carried into the complex $\omega$ plane together with the
poles that generate it. One should notice, however, that this is a cancellation and not a
suppression. The singularities of $\Pi_{B}$
sit at $\omega=\pm2E_{k}$, that is, at a distance $2\Gamma_{k}=4Nav/\omega$ from the real
axis, which tends to zero as $\omega$ grows, so that the curve of singularities approaches
the real axis asymptotically and the theory becomes effectively Hermitian in the
ultraviolet. Any mechanism that gives the excitations an ordinary width $\eta$ — an
interaction, a coupling to an environment — unbalances the pair, and the Hermitian
absorption is recovered once $\eta\gtrsim2\Gamma_{k}$. It is in this sense that the
Hermitian limit $av\to0$ is singular rather than smooth.

\section{The mixed channel}
\label{sec:A}

On the other hand, in $\Pi(m,m^{*})$ the two propagators carry conjugate masses, and the
relevant combinations are
\begin{equation}\label{eq:combs}
E_{1}-E_{2}=2i\Gamma_{k},\qquad E_{1}+E_{2}=2\varepsilon_{k},
\end{equation}
the first purely imaginary and the second real. The interband denominators are therefore
$\omega-2s\varepsilon_{k}$, and, since $\varepsilon_{k}\geq\varepsilon_{0}=N$, a cut opens
on the real axis at
\begin{equation}\label{eq:thresh}
\omega=2\varepsilon_{0}=2N,
\quad\text{that is}\quad
p^{2}=(m+m^{*})^{2}=4N^{2},
\end{equation}
which is precisely the threshold of the mixed pair found in \cite{LemesTedesco}, here
recovered as the optical gap of the broken channel. Carrying out the momentum integration
with the help of \eqref{eq:disp} we obtain the closed form
\begin{equation}\label{eq:sigmaA}
\sigma_{1}^{A}(\omega)=-\frac{k_{\omega}\,T_{+-}\,F}{2\pi\omega}
\left(\frac{\omega}{2}+\frac{8N^{2}a^{2}v^{2}}{\omega^{3}}\right),
\end{equation}
where $T_{+-}$ denotes the coefficient \eqref{eq:T} at $s=+$, $s'=-$ and
$(M_{1},M_{2})=(m,m^{*})$, while $\varepsilon=\omega/2$ and $\Gamma=2Nav/\omega$ are the
values selected by the threshold condition, so that
\begin{equation}\label{eq:sigmaAdata}
\begin{aligned}
k_{\omega}^{2}&=\frac{\omega^{2}}{4}-N^{2}+a^{2}v^{2}-\frac{4N^{2}a^{2}v^{2}}{\omega^{2}},\\
T_{+-}&=1+\frac{k_{\omega}^{2}/3+N^{2}+a^{2}v^{2}}
{\omega^{2}/4+4N^{2}a^{2}v^{2}/\omega^{2}},\\
F&={\rm Re}\big[n_{F}(\varepsilon+i\Gamma-\mu)-n_{F}(-\varepsilon+i\Gamma-\mu)\big].
\end{aligned}
\end{equation}
It is straightforward to confirm that $k_{\omega}$ vanishes at $\omega=2N$, as it must, and
that in the Hermitian limit $av\to0$ at vanishing temperature and chemical potential
\eqref{eq:sigmaA} reduces to twice the free Dirac result
$\frac{\omega}{12\pi}\big(1+2N^{2}/\omega^{2}\big)\sqrt{1-4N^{2}/\omega^{2}}$, the factor
two counting the two flavors; this is the check that fixes the overall sign in
\eqref{eq:Pi}. We have also checked \eqref{eq:sigmaA} against a direct numerical
evaluation of \eqref{eq:Pi}, with agreement to four significant figures.

The intraband sector of this channel deserves a separate comment, since it is the one in
which the width \eqref{eq:width} becomes visible. From \eqref{eq:combs} the four intraband
terms of the two orderings combine into
\begin{equation}\label{eq:intra}
\sum_{\rm intra}=-\frac{8\Gamma_{k}\big(\zeta_{k}+\bar{\zeta}_{k}\big)}{\omega^{2}+4\Gamma_{k}^{2}},
\quad
\begin{aligned}
\zeta_{k}&={\rm Im}\,n_{F}(\varepsilon_{k}+i\Gamma_{k}-\mu),\\
\bar{\zeta}_{k}&={\rm Im}\,n_{F}(-\varepsilon_{k}+i\Gamma_{k}-\mu),
\end{aligned}
\end{equation}
a Lorentzian of half-width $2\Gamma_{k}$ which, in the degenerate regime, is dominated by
the states at the Fermi surface and has the width $2\Gamma_{F}=2Nav/\mu$. This is the
shape of a Drude peak. However, the right-hand side of \eqref{eq:intra} is real, and
therefore contributes to the reactive part of the conductivity and not to $\sigma_{1}$.
The complex mass thus resolves the familiar ambiguity of the free gas at $\bq=0$, where
the intraband term is of the form $0/0$, and replaces it by a well-defined Lorentzian of
finite width, without generating any Joule heating: what one obtains is a Drude response
deprived of its dissipation.

\section{The response kernel in time}
\label{sec:time}

We proceed to answer the question which we regard as the more revealing one. The kernel of
the linear response, $J(t)=\int dt'\,\sigma(t-t')E(t')$, is obtained from
$\sigma(\omega)=iK(\omega)/(\omega+i0)$, with $K=\Pi_{\rm dia}+\Pi$ the full transverse
response. The constant $\Pi_{\rm dia}$ contributes the causal step
$\Pi_{\rm dia}\,\theta(t)$, and the whole of the non-trivial time dependence of the
intraband sector comes from \eqref{eq:intra}, whose poles sit at
$\omega=\pm2i\Gamma_{k}$. One of them lies in the upper half plane. Closing the contour in
each half plane one finds, for a single mode,
\begin{equation}\label{eq:kernel}
\frac{\sigma_{\rm intra}(t)}{\sigma_{\rm intra}(\infty)}=
\begin{cases}
1-\tfrac{1}{2}e^{-2\Gamma_{k}t}, & t>0,\\[4pt]
\tfrac{1}{2}e^{+2\Gamma_{k}t}, & t<0,
\end{cases}
\end{equation}
a continuous function which rises from zero at $t\to-\infty$ to unity at $t\to+\infty$ and
passes through one half at the instant at which the field is switched on. The response does
not vanish for negative times: a current is present before the field that drives it, over
a window
\begin{equation}\label{eq:precursor}
\tau_{\rm prec}=\frac{1}{2\Gamma_{F}}=\frac{\mu}{2Nav}.
\end{equation}
In the vacuum the intraband sector is absent, there being no carriers, and the precursor is
then generated by the branch point at $\omega=2N+2iav$ of Section \ref{sec:B}, which lies
in the upper half plane as well; the window becomes $1/2av$, fixed by the non-Hermitian
scale alone and independent of $\mu$ and $T$. The acausality is the one long known to
accompany complex conjugate poles \cite{LeeWick1969,Coleman1969}, here exhibited in a
transport coefficient rather than in a scattering amplitude, and it is the price of the
exact cancellation reported in Section \ref{sec:B}: a response function that is analytic in
the upper half plane and has no imaginary part on the real axis is necessarily a constant,
so that the vanishing of the absorptive part of Section \ref{sec:B} and the precursor of
\eqref{eq:kernel} are two readings of the same fact. One should add that
\eqref{eq:kernel} follows from the analytic continuation of the Euclidean correlator,
which is the definition adopted in \cite{LemesTedesco}, and that the paragraphs below
show this to be the wrong assignment of boundary value for a retarded response, so that
the precursor is an artifact of that assignment and not a property of the model. We note,
in contrast with the Lee--Wick case, that \eqref{eq:width} would make the window
\eqref{eq:precursor} grow with the energy of the mode, so that the acausality carried by
the common boundary value is a phenomenon of long times and low frequencies rather than a
short-distance disease.

Before the alternative, a word on what the choice of contour is doing, because the
precursor is a property of that choice and not of the model, and the alternative to it is
not innocuous either. Take a single intraband mode, with meromorphic coefficient
$K_{k}(z)=C_{k}/(z^{2}+a_{k}^{2})$, $a_{k}=2\Gamma_{k}>0$, and
$\sigma_{k}(z)=-iK_{k}(z)/z$, so that
$\sigma_{k}=i\sigma_{0k}\big(z^{-1}-z/(z^{2}+a_{k}^{2})\big)$ with
$\sigma_{0k}=-C_{k}/a_{k}^{2}$. Inverting along the real axis, with the pole at the origin
below the contour, returns \eqref{eq:kernel}: bounded, and noncausal. Inverting instead
along ${\rm Im}\,z=\gamma>a_{k}$, which encloses all three poles for $t>0$ and none for
$t<0$, returns
\begin{equation}\label{eq:causal-growing}
\frac{\sigma_{k}^{\rm causal}(t)}{\sigma_{0k}}=\theta(t)\,\big[1-\cosh(a_{k}t)\big],
\end{equation}
which is causal and unstable, with no finite late-time limit. The two differ by the
homogeneous growing solution,
\begin{equation}\label{eq:kernel-difference}
\frac{\sigma_{k}^{\rm axis}(t)-\sigma_{k}^{\rm causal}(t)}{\sigma_{0k}}
=\tfrac{1}{2}e^{a_{k}t},
\end{equation}
both satisfying $f''-a_{k}^{2}f=-a_{k}^{2}$ for $t>0$ but with
$f(0)=f'(0)=0$ in the causal case against $f(0)=1/2$, $f'(0)=a_{k}/2$ in the bounded one:
the bounded assignment cancels the future growth by admitting a past. The choice is
therefore between an advanced response and an exponentially growing one, and no local
contact term, being polynomial in frequency, can remove the offending pole. The growth is
a statement about the linearized mode and supplies no model of saturation.

With that said, the alternative for the diagonal channel. The cancellation of Section
\ref{sec:B} rests on assigning the same boundary value $\omega+i0$ to both terms of
\eqref{eq:assign}, although each has its singularities on a different side of the real
axis. Continuing each term instead from the half plane in which it is analytic replaces
$2\,{\rm Re}\,\Pi(m,m)$ by $2\Pi(m^{*},m^{*})$ on the real axis, and the absorptive part
no longer cancels: each interband denominator contributes a Lorentzian of width
$2\Gamma_{k}$, and
\begin{equation}\label{eq:LW}
\sigma_{1}^{B,{\rm ret}}(\omega)=\frac{2}{\omega}\,
{\rm Im}\,\Pi(m^{*},m^{*})(\omega)\;\neq\;0 .
\end{equation}
The two assignments do not differ by a small correction. They differ by the whole of the
Hermitian absorption, as Table \ref{tab:price} shows, and \eqref{eq:LW} carries the
ordinary ultraviolet behaviour of the massless theory, $\sigma_{1}\to\omega/6\pi$ with the
factor counting the two flavors: we find $\sigma_{1}^{B,{\rm ret}}/\omega=0.05304$ at
$\omega=100\,N$ for $av=0.40\,N$, against $1/6\pi=0.05305$, while the common boundary
value gives zero at every frequency however large.

It should be said plainly that this second assignment does not produce a retarded
function either. The interband sum contains both signs of frequency, and with the
surviving mass $m^{*}$ the two poles are
\begin{equation}\label{eq:bothsigns}
\omega=+2\varepsilon_{k}-2i\Gamma_{k},
\qquad
\omega=-2\varepsilon_{k}+2i\Gamma_{k},
\end{equation}
the second of which lies in the upper half plane for every $\bk$. Moving the
positive-frequency singularity below the axis therefore does not move them all, and any
prescription meant to retain only damped singularities would have to specify the complete
function at both signs of frequency and say how it is related to the Euclidean
correlator. We impose none here. The contest is thus not between a good assignment and a
bad one; neither is a stable retarded response, and Section \ref{sec:euclid} shows that
the question is in any case prior to the choice.

One would want to settle the matter with the optical sum rule, and it cannot be done, for
a reason that Section \ref{sec:euclid} establishes as a theorem rather than an
expectation. The rule in its familiar form, and with it the requirement that a deformation
by a mass leave the ultraviolet asymptotics alone, is a statement about a K\"all\'en--%
Lehmann representation with a \emph{positive measure}. The diagonal channel has none: its
Euclidean correlator \eqref{eq:CB} changes sign in $\tau$, so by Osterwalder--Schrader it
admits no positive-norm realization, and we exhibit both the test function of negative
norm and the order \eqref{eq:norder} at which the moment sequence first fails. Every
repair one might attempt is closed in Table \ref{tab:escapes}, the choice of boundary
value among them --- that test is performed in the Euclidean region and never continues to
real $\omega$ at all.

We are careful about what this does and does not say. The current $J_{B}$ does reach a
state of $\eta$-norm $+1$, the symmetrized combination of its plane, and that norm is
conserved exactly by \eqref{eq:etacons}. But the $\eta$-form and the
Osterwalder--Schrader form are different pairings, and it is the second that decides
whether a Hilbert space exists; a positive $\eta$-norm on one state buys nothing. The
vanishing of \eqref{eq:sigmaB} and the precursor of \eqref{eq:kernel} are therefore
statements about a correlator with no positive spectral representation, and the choice of
boundary value for $\Pi_{B}$ is a choice about such a correlator. Table \ref{tab:price}
retains its full interest as the measure of what the two assignments differ by. We regard
the question as disposed of rather than answered, and prefer to say so.

The physical content sits in the mixed channel, and there the discussion cannot reach,
for two independent reasons. The first is kinematic: the interband singularities sit at
$\omega=2\varepsilon_{k}$, on the real axis, so that no choice of sheet enters, and we
have checked numerically that the intraband sector, the only part of $\Pi_{A}$ with
singularities off the axis, contributes nothing to ${\rm Im}\,\Pi_{A}$. The second is
exact: by \eqref{eq:JZ2} the two currents have opposite $\mathbb{Z}_{2}$ parity, so that
$\langle J_{A}J_{B}\rangle=0$ and no prescription adopted for one channel can feed into
the other. Everything obtained in Sections \ref{sec:A}, \ref{sec:time} and \ref{sec:pos}
is therefore independent of the present discussion, by symmetry and not only by
inspection. Appendix \ref{app:cuts} reaches the same separation from the analytic
continuation itself: the channel in which the ambiguity arises is the one whose cut is
unbalanced, and which for that reason admits no particle representation under any
prescription, while the channel that does admit one has its singularities on the real
axis.

One further question belongs here, and Section \ref{sec:pos} answers it: where the
ultraviolet weight of the theory actually is. Doing the vacuum integral of the mixed
channel in closed form shows that the whole of the free two-flavor asymptotics
$\omega/6\pi$ is carried by that channel, so that the weight Table \ref{tab:price}
measures is not missing from the model but sits where no prescription is in question.

\begin{table}[b]
\caption{The two assignments of boundary value, in units $N=1$, for $T=0.15\,N$,
$\mu=1.4\,N$ and $av=0.40\,N$. The second column gives both terms the boundary value
$\omega+i0$ and vanishes identically by \eqref{eq:sigmaB}; the third continues each term
from its own half plane, Eq. \eqref{eq:LW}; the fourth is the same integral at $av\to0$,
that is the ordinary Hermitian absorption at the same temperature and density, which the
third column approaches in the ultraviolet.}
\label{tab:price}
\begin{ruledtabular}
\begin{tabular}{cccc}
$\omega/N$ & $\sigma_{1}^{B}$ & $\sigma_{1}^{B,{\rm ret}}$ & $\sigma_{1}^{B}|_{av\to0}$ \\
\colrule
$2.5$ & $0$ & $0.0299$ & $0.0283$ \\
$3.0$ & $0$ & $0.0963$ & $0.0958$ \\
$4.0$ & $0$ & $0.1949$ & $0.2030$ \\
$6.0$ & $0$ & $0.3083$ & $0.3168$ \\
$9.0$ & $0$ & $0.4710$ & $0.4770$ \\
$45.0$ & $0$ & $2.3860$ & $2.3873$ \\
$100.0$ & $0$ & $5.3043$ & $5.3052$ \\
\end{tabular}
\end{ruledtabular}
\end{table}

\begin{figure*}[t]
\centering
\IfFileExists{fig_conductivity.png}
 {\includegraphics[width=\textwidth]{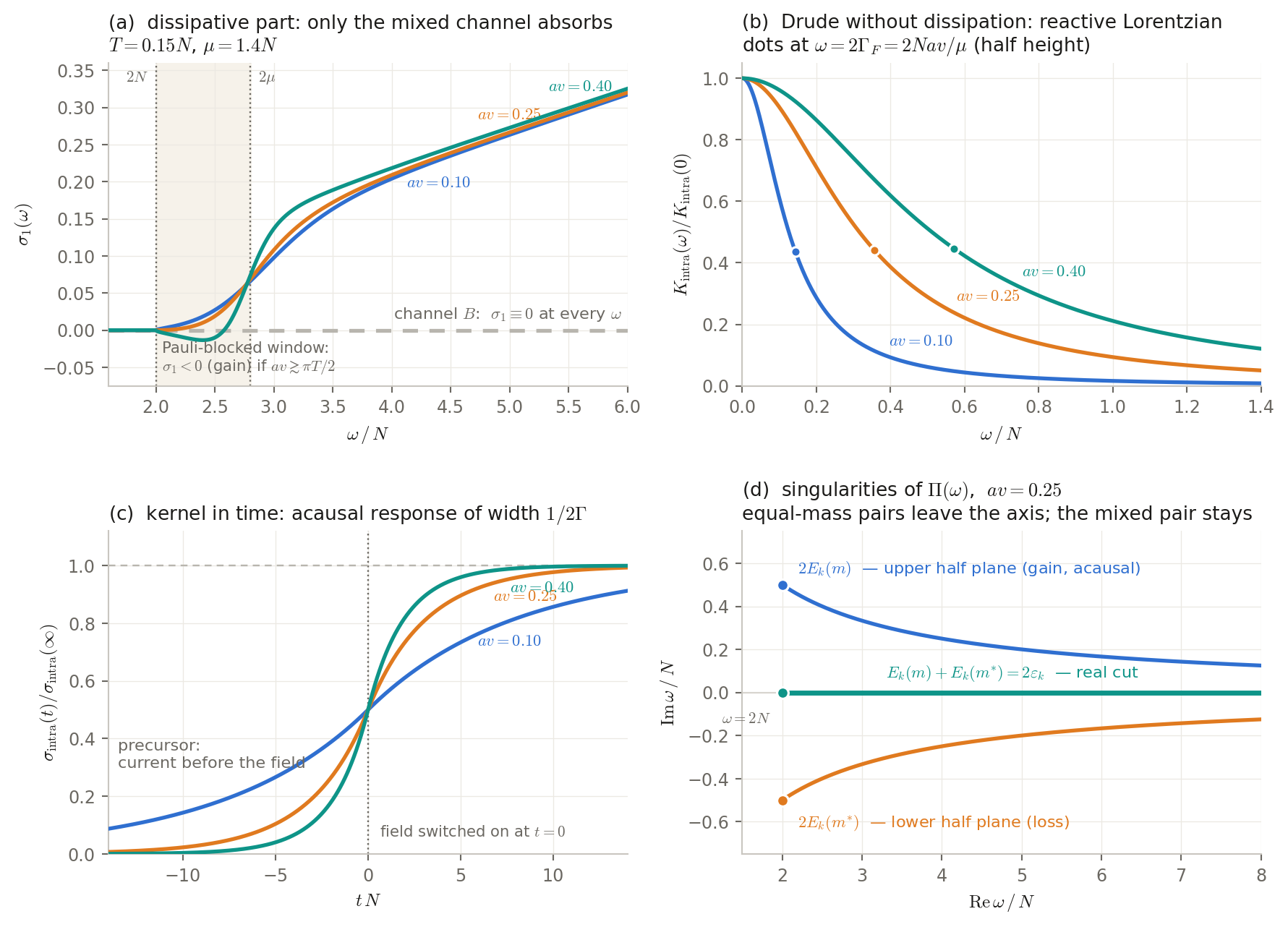}}
 {\framebox[\textwidth][c]{\rule{0pt}{6cm}\texttt{fig\_conductivity.png}}}
\caption{Linear response of the complex-mass fermions, in units $N=1$, for
$T=0.15\,N$ and $\mu=1.4\,N$. (a) Dissipative conductivity: the unbroken channel vanishes
identically, while the mixed channel absorbs above $\omega=2N$, with the Pauli-blocked
window $2N<\omega<2\mu$ shaded; the excursion to negative values inside that window is the
gain discussed in Section \ref{sec:pos}. (b) The intraband Lorentzian of the mixed
channel, normalized at the origin, with the half-height points at
$\omega=2\Gamma_{F}=2Nav/\mu$. (c) The response kernel \eqref{eq:kernel}, showing the
precursor at negative times. (d) The singularities of the polarization in the complex
$\omega$ plane: the equal-mass pairs leave the real axis in opposite directions, while the
mixed pair remains on it and opens the cut at $\omega=2N$.}
\label{fig:cond}
\end{figure*}

\section{Positivity of the spectral density}
\label{sec:pos}

It remains to examine the sign of the spectral density, a question that the scalar
computation of \cite{LemesTedesco} left open.

\subsection{The vacuum}

For $T\to0$ and $\mu=0$ the occupation factor of \eqref{eq:sigmaAdata} reduces to $F=-1$,
and the residue $T_{+-}$ of \eqref{eq:T} is
\begin{equation}\label{eq:Tpm}
T_{+-}=1+\frac{\bk^{2}/3+|m|^{2}}{|E_{k}|^{2}}>0,
\end{equation}
strictly positive, since $E_{1}E_{2}=|\bk^{2}+m^{2}|$ is real and positive for real $\bk$.
Then
\begin{equation}\label{eq:pos}
\sigma_{1}^{A}(\omega)=\frac{2\pi}{\omega}\int\!\frac{d^{3}k}{(2\pi)^{3}}\,
T_{+-}(\bk)\,\delta(\omega-2\varepsilon_{k})\;>\;0,
\end{equation}
so that in the vector channel the transverse spectral density is not merely real but
positive, and manifestly so. To our knowledge this is the first positivity statement
available for a composite correlator of this model; Section \ref{sec:euclid} will promote
it from a computed outcome to a theorem, by showing that the Euclidean correlator of this
channel is completely monotone. The qualifier is not decorative: at $\bq=0$ the
conductivity is a spatial response and probes the transverse form factor alone, and we
shall find below that the longitudinal one behaves differently.

The delta function can be removed and the integral done in closed form. With
$D_{k}=|E_{k}|^{2}=\varepsilon_{k}^{2}+\Gamma_{k}^{2}$ and
$\varepsilon_{k}\Gamma_{k}=Nav$, Appendix \ref{app:disp} gives
\begin{equation}\label{eq:kofeps}
\bk^{2}=\frac{(\varepsilon_{k}^{2}-N^{2})(\varepsilon_{k}^{2}+a^{2}v^{2})}
{\varepsilon_{k}^{2}},
\qquad
k\,\frac{dk}{d\varepsilon_{k}}=\frac{D_{k}}{\varepsilon_{k}},
\end{equation}
so that in the variable $s=4\varepsilon_{k}^{2}$, with
\begin{equation}\label{eq:lambda}
\lambda(s)=(s-4N^{2})(s+4a^{2}v^{2}),
\qquad k=\frac{\sqrt{\lambda(s)}}{2\sqrt{s}},
\end{equation}
one has $d^{3}k/(2\pi)^{3}=(2\pi^{2})^{-1}k\,(D_{k}/\varepsilon_{k})\,d\varepsilon_{k}$.
The residue \eqref{eq:Tpm} collapses algebraically: writing out $D_{k}$ and $\bk^{2}$ in
terms of $\varepsilon_{k}$,
\begin{equation}\label{eq:DT1}
\begin{aligned}
D_{k}T_{+-}&=D_{k}+\tfrac{1}{3}\bk^{2}+|m|^{2}\\
&=\tfrac{2}{3}\big(2\varepsilon_{k}^{2}+N^{2}\big)
\Big(1+\frac{a^{2}v^{2}}{\varepsilon_{k}^{2}}\Big)\\
&=\frac{(s+2N^{2})(s+4a^{2}v^{2})}{3s},
\end{aligned}
\end{equation}
all four terms combining into a product. Evaluating \eqref{eq:pos} at
$\varepsilon_{k}=\omega/2$ therefore gives, for $\omega>2N$,
\begin{equation}\label{eq:sigmaclosed}
\sigma_{1}^{A}(\omega)=\frac{\sqrt{\lambda(\omega^{2})}\,
(\omega^{2}+2N^{2})(\omega^{2}+4a^{2}v^{2})}{6\pi\,\omega^{5}},
\end{equation}
and zero below. Two limits check it. At $av\to0$ it returns
$(\omega/6\pi)(1+2N^{2}/\omega^{2})\sqrt{1-4N^{2}/\omega^{2}}$, which is twice the free
Dirac result, the factor counting the two flavors; and, since
$\rho_{1}(s)=s/6\pi^{2}+a^{2}v^{2}/\pi^{2}+O(s^{-1})$ in the notation of \eqref{eq:rhos}
below,
\begin{equation}\label{eq:UVA}
\sigma_{1}^{A}(\omega)=\frac{\omega}{6\pi}+\frac{a^{2}v^{2}}{\pi\omega}+O(\omega^{-3}).
\end{equation}
This settles the question raised in Section \ref{sec:time}. The leading term is exactly
the free value for the two flavors, the same number that \eqref{eq:LW} reproduces for the
diagonal channel under the second assignment of boundary value. The ultraviolet weight
that Table \ref{tab:price} measures is therefore not missing from the model: it is carried
in full by the mixed channel, whose singularities lie on the real axis and whose
prescription is not in question. What the cancellation of \eqref{eq:sigmaB} says is the
narrower thing, that the \emph{massless} probe does not see it.

The same reduction delivers the other form factor, and there the sign is the other one.
The covariant vacuum calculation of Appendix \ref{app:spectral} decomposes the mixed
correlator into a transverse and a longitudinal part, with spectral densities
\begin{equation}\label{eq:rhos}
\begin{aligned}
\rho_{1}(s)&=\frac{\sqrt{\lambda(s)}\,(s+2N^{2})(s+4a^{2}v^{2})}{6\pi^{2}s^{2}},\\
\rho_{0}(s)&=-\frac{a^{2}v^{2}\sqrt{\lambda(s)}\,(s-4N^{2})}{\pi^{2}s^{2}},
\end{aligned}
\end{equation}
both supported on $s\geq4N^{2}$, the first being \eqref{eq:sigmaclosed} again through
$\sigma_{1}^{A}(\omega)=(\pi/\omega)\rho_{1}(\omega^{2})$. The longitudinal one is
present because $J_{A}$ is not conserved, its divergence being proportional to $m-m^{*}$
as noted after \eqref{eq:currents}, and it is \emph{negative}. That this is the
non-Hermiticity and not a convention follows from an identity. For any two masses, with
$E_{i}=\sqrt{\bk^{2}+m_{i}^{2}}$,
\begin{equation}\label{eq:lemma}
(E_{1}E_{2})^{2}-\big(\bk^{2}+m_{1}m_{2}\big)^{2}=\bk^{2}(m_{1}-m_{2})^{2},
\end{equation}
as one sees by expanding both sides. The longitudinal residue is
$T_{0}=1-(\bk^{2}+|m|^{2})/D_{k}$, whose sign is that of
$E_{1}E_{2}-\bk^{2}-m_{1}m_{2}$ at $m_{1}=m$, $m_{2}=m^{*}$. For two \emph{real} masses
the right-hand side of \eqref{eq:lemma} is non-negative, so that $T_{0}\geq0$, with
equality precisely at $m_{1}=m_{2}$, where the current is conserved and there is no
longitudinal part at all. For the conjugate pair $(m-m^{*})^{2}=-4a^{2}v^{2}$, both
$E_{1}E_{2}=D_{k}$ and $\bk^{2}+|m|^{2}$ are real and positive, and
\begin{equation}\label{eq:gap}
D_{k}^{2}-\big(\bk^{2}+|m|^{2}\big)^{2}=-4a^{2}v^{2}\bk^{2}<0,
\end{equation}
so $D_{k}<\bk^{2}+|m|^{2}$ and $T_{0}<0$ for every $\bk\neq0$. Explicitly
$D_{k}T_{0}=-2a^{2}v^{2}(s-4N^{2})/s$, which is the second line of \eqref{eq:rhos}. The
negative longitudinal weight is thus the square of the mass splitting, and the splitting
is imaginary.

Two things follow. The first is that the positivity of \eqref{eq:pos} is a statement about
the transverse channel, which is the one the conductivity probes: at $\bq=0$ the response
is spatial and $\rho_{0}$ does not enter. The second is that the current correlator as a
whole is not positive, since in the rest frame its spectral tensor is
${\rm diag}(\rho_{0},\rho_{1},\rho_{1},\rho_{1})$. Section \ref{sec:composite} shows that
the offending component is removed by passing from the current to the antisymmetric
derivative of it, and that what remains is positive, causal, and carried by an honest
Hilbert space.

What secures \eqref{eq:pos}, at the level of the computation, is that the residue of the
mixed pair is a sum of positive terms, the non-Hermiticity having been confined to the
position of the pole and not to its weight; what secures it structurally is
\eqref{eq:pairnorm}, the intermediate state of this channel being the symmetrized
conjugate pair, of norm $+1$, to which the current couples while remaining
$\eta$-orthogonal to the partner of norm $-1$. The sign in \eqref{eq:pos} is in that sense
not an accident of the Dirac algebra but the norm of the $i$-particle pair seen through
the current. It should be said that this is an explanation of a computed result and not a
substitute for the computation: the same plane structure is present in the diagonal
channel, where no positivity statement is available, and what distinguishes the two is the
position of the singularities and the $\mathbb{Z}_{2}$ parity of \eqref{eq:JZ2}, not the
signature of the metric. Nor does the argument from \eqref{eq:pairnorm} extend to
$\rho_{0}$: it concerns the flavor structure of the bilinear, and \eqref{eq:gap} is a
statement about its Lorentz structure, which the one- and two-particle norms of Section
\ref{sec:metric} do not see.

\subsection{Finite temperature: a sharp bound}
\label{sec:bound}

In a medium the sign of $F$ is no longer fixed, and the question becomes one about the
continuation of the occupation numbers, which Section \ref{sec:exact} has placed on a firm
footing. Writing
\begin{equation}\label{eq:tc}
t=e^{(\varepsilon-\mu)/T},\qquad c=\cos(\Gamma/T),
\end{equation}
a short computation gives the closed form
\begin{equation}\label{eq:ReNF}
{\rm Re}\,n_{F}(\varepsilon+i\Gamma-\mu)=\frac{ct+1}{t^{2}+2ct+1},
\end{equation}
whose denominator is strictly positive for $|c|<1$, and whose derivative is
\begin{equation}\label{eq:dReNF}
\frac{\partial}{\partial\varepsilon}{\rm Re}\,n_{F}
=-\frac{t\,\big(ct^{2}+2t+c\big)}{T\,\big(t^{2}+2ct+1\big)^{2}} .
\end{equation}
For $c>0$ the bracket in \eqref{eq:dReNF} is strictly positive on $t>0$, so that
${\rm Re}\,n_{F}$ is a strictly decreasing function of $\varepsilon$ at fixed $\Gamma$.
Since the two occupation factors in $F$ are evaluated at the same $\Gamma$ and at
$\varepsilon$ and $-\varepsilon$, this gives $F<0$ and therefore
$\sigma_{1}^{A}(\omega)>0$. Population inversion thus requires $c<0$, that is
$\Gamma>\pi T/2$, and with the bound \eqref{eq:width} we arrive at
\begin{equation}\label{eq:bound}
av<\frac{\pi T}{2}
\qquad\Longrightarrow\qquad
\sigma_{1}^{A}(\omega)>0\ \ \text{for every }\omega,
\end{equation}
a positivity statement at finite temperature and density with an explicit bound, the
non-Hermitian scale being measured against the first Matsubara frequency. The bound is
sharp: a numerical determination of the true onset $av_{c}$ of negative
$\sigma_{1}^{A}$, scanning $T\in[0.08,0.35]\,N$ and $\mu\in[1.1,4.2]\,N$, never falls
below $\pi T/2$ and saturates it to three decimal places as $\mu/N$ grows, as shown in
Fig.~\ref{fig:bound}.

\subsection{Above the bound the medium amplifies}

\begin{figure*}[t]
\centering
\IfFileExists{fig_positivity_bound.png}
 {\includegraphics[width=\textwidth]{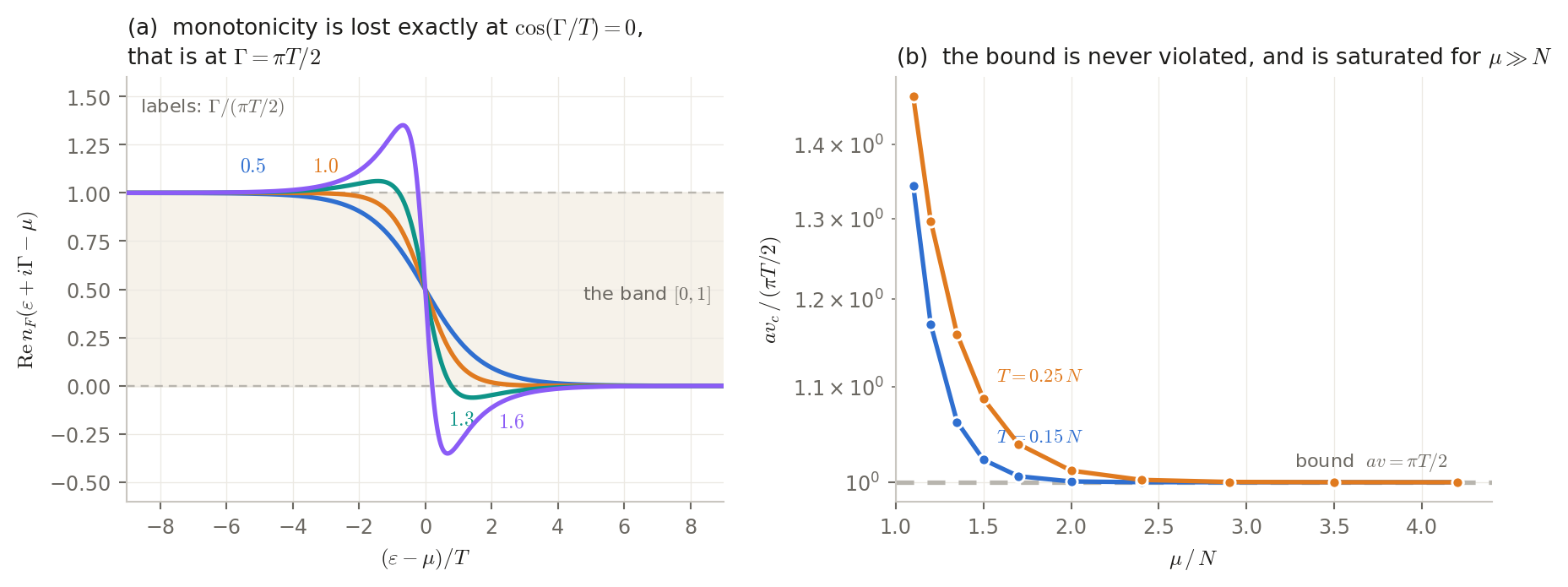}}
 {\framebox[\textwidth][c]{\rule{0pt}{6cm}\texttt{fig\_positivity\_bound.png}}}
\caption{The positivity bound \eqref{eq:bound}. (a) The real part of the occupation
number \eqref{eq:ReNF} as a function of $(\varepsilon-\mu)/T$, for several values of
$\Gamma$ in units of $\pi T/2$. Monotonicity, and with it $F<0$ and
$\sigma_{1}^{A}>0$, is lost exactly when $\cos(\Gamma/T)$ changes sign, and the occupation
leaves the interval $[0,1]$ at the same point. (b) The numerically determined onset
$av_{c}$ of negative $\sigma_{1}^{A}$, in units of the bound, as a function of $\mu/N$ for
two temperatures: the bound is never violated and is saturated for $\mu\gg N$.}
\label{fig:bound}
\end{figure*}

For $c<0$ the bracket of \eqref{eq:dReNF} vanishes at
$t_{\mp}=\big(1\mp\sqrt{1-c^{2}}\,\big)/|c|$, and ${\rm Re}\,n_{F}$ increases with
$\varepsilon$ for $t<t_{-}$ and for $t>t_{+}$, that is in the tails far from the chemical
potential. The effective occupation of the upper state may then exceed that of the lower
one and $\sigma_{1}^{A}$ turns negative: the medium amplifies instead of absorbing. Figure
\ref{fig:cond}(a) shows the effect inside the Pauli-blocked window $2N<\omega<2\mu$, which
is where it first appears, and Fig.~\ref{fig:bound}(a) shows the mechanism, the real part
of the occupation number leaving the interval $[0,1]$ as soon as $\Gamma$ exceeds
$\pi T/2$.

The onset can be located exactly. Away from a Matsubara collision the denominator of
\eqref{eq:dReNF} is positive, so the continued interband coefficient is negative precisely
when
\begin{equation}\label{eq:negative-condition}
\cos(\Gamma/T)<-\frac{\cosh(\varepsilon/T)}{\cosh(\mu/T)} ,
\end{equation}
which requires $|\mu|>\varepsilon\geq N$: for $|\mu|\leq N$ no negative coefficient occurs
at any $av$ away from the singular loci. Solving \eqref{eq:negative-condition} at
equality gives
$av=(\varepsilon T/N)\arccos[-\cosh(\varepsilon/T)/\cosh(|\mu|/T)]$, and both factors
increase with $\varepsilon$ on $N\leq\varepsilon<|\mu|$, so the infimum sits at the
threshold $\varepsilon=N$:
\begin{equation}\label{eq:avc}
(av)_{c}(T,\mu)=T\arccos\!\left[-\frac{\cosh(N/T)}{\cosh(|\mu|/T)}\right].
\end{equation}
The spectral weight vanishes at that endpoint, but for $av$ just above $(av)_{c}$ an open
interval above threshold satisfies \eqref{eq:negative-condition}, and we find the onset
numerically at $(av)_{c}$ to five figures. As $|\mu|/T\to\infty$ the argument of the
arccosine tends to zero and $(av)_{c}\to\pi T/2$, which proves that the bound
\eqref{eq:bound} is uniform in the chemical potential and that it is saturated, not merely
respected. At larger $av$ the cosine produces further sign intervals, so
\eqref{eq:avc} describes the first onset and not a monotonic phase boundary.

This is a novelty, as, up to now, the discussion of complex conjugate poles has been
carried out in the vacuum, where the conjugation pairing is enough to secure reality and,
as \eqref{eq:pos} shows, positivity as well. One should notice that an occupation number
outside $[0,1]$ is not in conflict with the exclusion principle here, since by
\eqref{eq:norms} the diagonal modes of complex mass have vanishing $\eta$-norm and are
not states whose occupation is required to be a probability; the quantities that carry
norm are the conjugate pairings, and it is their difference, through $F$, that decides
between absorption and gain.

\section{The Euclidean criterion}
\label{sec:euclid}

The two results so far --- the positivity of \eqref{eq:pos} and the vanishing of
\eqref{eq:sigmaB} --- are statements on the real frequency axis, and each was argued with
a boundary value already chosen. There is a test that needs none, and it settles both
channels at once. At vanishing spatial momentum, performing the frequency integral of the
one-loop bubble and transforming back in $Q_{4}$ gives, for $\tau>0$ and with the two
orderings summed,
\begin{equation}\label{eq:Ctau}
C(\tau)=\!\int\!\frac{d^{3}k}{(2\pi)^{3}}\,2\,{\rm Re}
\Big\{\Big[1+\frac{\bk^{2}/3+m_{1}m_{2}}{E_{1}E_{2}}\Big]
e^{-(E_{1}+E_{2})\tau}\Big\},
\end{equation}
$E_{i}=\sqrt{\bk^{2}+m_{i}^{2}}$ on the principal branch. Everything turns on whether
$E_{1}+E_{2}$ is real.

The criterion is the Osterwalder--Schrader one \cite{OS1973}. A field with a positive-norm
realization has a reflection-positive Euclidean correlator, and at zero spatial momentum
this reduces to the statement that $\tau\mapsto C(\tau)$ is completely monotone,
$(-1)^{n}C^{(n)}(\tau)\geq0$ for all $n$ and all $\tau>0$; Bernstein's theorem then
returns a positive measure, which is a K\"all\'en--Lehmann representation with
$\rho\geq0$. Nothing in this depends on how one later continues to real $\omega$.

\subsection{The mixed channel: positivity as a theorem}

With $m_{1}=m$ and $m_{2}=m^{*}$ one has $E_{1}E_{2}=|E_{k}|^{2}=D_{k}$ and
$E_{1}+E_{2}=2\varepsilon_{k}$, both real and positive, the bracket in \eqref{eq:Ctau} is
the residue $T_{+-}$ of \eqref{eq:Tpm}, and the real part is taken of something already
real:
\begin{equation}\label{eq:CA}
C_{A}(\tau)=\int\!\frac{d^{3}k}{(2\pi)^{3}}\,T_{+-}(\bk)\,e^{-2\varepsilon_{k}\tau}.
\end{equation}
This is a positive superposition of decaying exponentials, so that
\begin{equation}\label{eq:CAmono}
(-1)^{n}C_{A}^{(n)}(\tau)=\!\int\!\frac{d^{3}k}{(2\pi)^{3}}
(2\varepsilon_{k})^{n}T_{+-}(\bk)e^{-2\varepsilon_{k}\tau}>0
\end{equation}
for every $n$ and every $\tau>0$, each factor being separately positive. $C_{A}$ is
completely monotone, and Bernstein gives
\begin{equation}\label{eq:rho1pos}
\rho_{1}(s)\geq0 .
\end{equation}
The positivity of the mixed channel is therefore a theorem and not a computed outcome, and
the proof uses only two facts: that the residue $T_{+-}$ is positive, and that the
conjugate pair has a real sum of energies. The closed form \eqref{eq:sigmaclosed} is then
a corollary one may check, as we have.

\subsection{The diagonal channel carries negative norm}

With $m_{1}=m_{2}=m$ the sum of energies is $E_{1}+E_{2}=2E_{k}=2(\varepsilon_{k}
+i\Gamma_{k})$, which is not real, and \eqref{eq:Ctau} becomes
\begin{equation}\label{eq:CB}
C_{B}(\tau)=\int\!\frac{d^{3}k}{(2\pi)^{3}}\,2|W_{k}|\,e^{-2\varepsilon_{k}\tau}
\cos\big(2\Gamma_{k}\tau+\varphi_{k}\big),
\end{equation}
$W_{k}=|W_{k}|e^{i\varphi_{k}}=1+(\bk^{2}/3+m^{2})/E_{k}^{2}$. A completely monotone
function is positive and cannot change sign. This one does. For $N=1$ and $av=0.6$,
$C_{B}=1.82\times10^{-2}$ at $\tau=1$, $-1.16\times10^{-4}$ at $\tau=2$ and
$+2.44\times10^{-7}$ at $\tau=5$: it oscillates, with the first zero near
$\tau\simeq1.88$, and the zero tracks the quarter period $\pi/4av$ of the slowest mode,
$\Gamma_{k}\leq\Gamma_{0}=av$ by \eqref{eq:width}. Any test function supported where
$C_{B}<0$ is a vector of negative Osterwalder--Schrader norm.

The same conclusion follows from the Euclidean form factor alone, with no reference to
$\tau$. If $A(x)=\int\rho(s)ds/(s+x)$ with $\rho\geq0$ then
$A''(x)=2\int\rho(s)(s+x)^{-3}ds$, so that
\begin{equation}\label{eq:moments}
c_{n}=\frac{(-1)^{n}A''^{(n)}(x_{0})}{(n+2)!}
=\int\frac{\rho(s)\,ds}{(s+x_{0})^{n+3}}
\end{equation}
is a Hausdorff moment sequence of a positive measure, and the Hankel matrix
$H_{jk}=c_{j+k}$ must be positive semidefinite. Using \eqref{eq:ddparam} with
$\mathcal{A}=|m|^{2}=m^{2}$ for the diagonal pair, the kernel is
$(m^{2}+u\bar u\,x)^{-(n+1)}$ with a \emph{complex} denominator, whose real part carries
$\cos\big[(n+1)\arg(m^{2}+u\bar u\,x)\big]$ and changes sign at
\begin{equation}\label{eq:norder}
n^{*}\simeq\frac{\pi}{2\arg m^{2}}=\frac{\pi}{4\arctan(av/N)} .
\end{equation}
Measured against predicted, the first $n$ with $c_{n}<0$ is $4$ against $4.0$ at
$av=0.2\,N$, $2$ against $2.1$ at $0.4\,N$, $1$ against $1.5$ at $0.6\,N$ and $3$ against
$3.2$ for $N=2$, $av=0.7$. The Hankel matrix has a negative eigenvalue of relative size
$0.24$ to $0.58$ over the same range, stable in its dimension, and its eigenvector is an
explicit witness. In the mixed channel the same test is passed to the limit of double
precision, as \eqref{eq:CAmono} requires.

\subsection{Nothing repairs it}

The obstruction is not an artifact of a choice, and it is worth closing the routes one by
one; Table \ref{tab:escapes} collects them.

\begin{table}[b]
\caption{The ways out of \eqref{eq:CB}, and why each is closed.}
\label{tab:escapes}
\begin{ruledtabular}
\begin{tabular}{ll}
proposed repair & why it fails \\
\colrule
another boundary value & the test is Euclidean \\
contact terms & $A''$ is blind to them \\
taking the curl & weight already transverse \\
mixing with $J_{A}$ & $\mathbb{Z}_{2}$ blocks the form \\
$\hat\Theta$ or $\hat\Theta^{C}$ & $J_{B}$ is even under both \\
nonzero $\bq$ & $\bq=0$ already fails \\
a regime of $av$ & only $av=0$ \\
filtering a subspace & the positive cone is unstable \\
a BRST quartet & $J_{B}$ is not $s$-exact \\
\end{tabular}
\end{ruledtabular}
\end{table}

A different boundary value cannot help, because \eqref{eq:CB} and \eqref{eq:moments} are
computed from the Euclidean correlator and from \eqref{eq:ddparam}, where
${\rm Re}(m^{2}+u\bar u\,x)>0$ fixes the branch and no continuation to real $\omega$ is
ever performed. Contact terms cannot help either: $[A]=2$, so renormalizability admits
only $A(0)+xA'(0)$, a polynomial of degree one, which $A''$ annihilates; every $c_{n}$ is
untouched, and $n^{*}\geq1$. The curl, which removed the negative weight of the mixed
channel in Section \ref{sec:composite}, does nothing here, because $J_{B}$ is conserved:
its longitudinal form factor vanishes identically, by \eqref{eq:lemma} with
$m_{1}=m_{2}$, so the whole of its weight is already transverse and $BP_{1}B^{\dagger}$
inherits it, reweighted by $s>0$. Mixing with $J_{A}$ is forbidden by the symmetry that
protects $J_{A}$: by \eqref{eq:JZ2} the two currents have opposite $\mathbb{Z}_{2}$
parity, $\langle J_{A}J_{B}\rangle=0$, and the Gram form of
$\alpha J_{A}+\beta J_{B}$ is $|\alpha|^{2}G_{A}+|\beta|^{2}G_{B}$, block diagonal, so a
negative direction of $G_{B}$ survives for every $\beta\neq0$. The antilinear $\Theta$ of
\eqref{eq:Theta} and the exchange $\mathcal{C}$ of \eqref{eq:Cdef} both leave $J_{B}$
invariant, so neither $\hat\Theta=\Theta_{\rm OS}\circ\Theta$ nor
$\hat\Theta^{C}=\Theta_{\rm OS}\circ\mathcal{C}\circ\Theta$ differs from the plain
reflection on this operator; and the $\mathcal{C}$-odd partner
$\overline{\Psi}_{m}\gamma\Psi_{m}-\overline{\Psi}_{m^{*}}\gamma\Psi_{m^{*}}$ has the
\emph{same} two-point function at one loop, the cross term requiring a mixed propagator
and vanishing. A single negative direction at $\bq=0$ is already fatal, so no spatial
momentum helps. Finally, $n^{*}$ in \eqref{eq:norder} is finite for every $av>0$; as
$av\to0$ it runs to infinity --- $n^{*}=17$ at $av=0.05\,N$, $n^{*}=88$ at $0.01\,N$ ---
and the negative eigenvalue goes to zero with it. The obstruction is the non-Hermiticity
itself, switched on continuously.

One line explains why the list is exhaustive. Reflection positivity at $\bq=0$ asks that
$C(\tau_{a}+\tau_{b})$ be of the form $\sum_{n}\overline{f_{n}(\tau_{a})}f_{n}(\tau_{b})$,
that is, that the two time arguments carry conjugate exponents. In \eqref{eq:Ctau} the
exponent is $E_{1}+E_{2}$, and it is self-conjugate exactly when the loop already carries
one line of mass $m$ and one of mass $m^{*}$ --- when the cut is \emph{balanced} in the
sense of Appendix \ref{app:cuts}. The balanced-cut criterion and reflection positivity are
one condition seen twice, and the diagonal channel fails both by construction.

\subsection{Can a positive subspace be filtered out?}
\label{sec:filter}

The natural reaction to an indefinite form is to look for a subspace on which it is
positive, and then to quotient: this is what Gupta--Bleuler does in electrodynamics and
what the quartet mechanism of Kugo and Ojima does in a covariant gauge
\cite{Piguet:1995er}. The question is a sharp one here, and it has an answer in each
direction.

A subspace of \emph{states} is not merely a linear subspace. In the
Osterwalder--Schrader construction the Euclidean time translation $e^{-sH}$ acts on test
functions by shifting their support, so a candidate physical subspace must be stable under
that semigroup; otherwise there is no Hamiltonian defined on it and no reconstruction. Let
$|f\rangle$ be built from a test function $\sum_{a}c_{a}\delta(\tau-\tau_{a})$ with
$\tau_{a}>0$. The norm of its evolution is
\begin{equation}\label{eq:Qs}
Q(s)=\sum_{a,b}c_{a}c_{b}\,C(\tau_{a}+\tau_{b}+2s),
\end{equation}
and both exponentials in \eqref{eq:Ctau} factorize through it. In the mixed channel this
gives
\begin{equation}\label{eq:QA}
Q_{A}(s)=\!\int\!\frac{d^{3}k}{(2\pi)^{3}}T_{+-}(\bk)
\Big[\sum_{a}c_{a}e^{-2\varepsilon_{k}\tau_{a}}\Big]^{2}e^{-4\varepsilon_{k}s}\;\geq\;0,
\end{equation}
a manifest square with a positive weight, for every vector and every $s$: the positive
cone is stable, and the reconstruction goes through. In the diagonal channel the same
step gives
\begin{equation}\label{eq:QB}
Q_{B}(s)=\!\int\!\frac{d^{3}k}{(2\pi)^{3}}2\,{\rm Re}\Big\{W_{k}
\Big[\sum_{a}c_{a}e^{-2E_{k}\tau_{a}}\Big]^{2}e^{-4E_{k}s}\Big\},
\end{equation}
a \emph{complex} square. As $s$ grows the integral is governed by the bottom of the band,
$\varepsilon_{k}\to N$ and $\Gamma_{k}\to av$, so that
\begin{equation}\label{eq:Qasym}
Q_{B}(s)\;\sim\;|Z|\,e^{-4Ns}\cos\!\big(4av\,s+\phi\big),
\end{equation}
which changes sign; the bracket is a finite sum of exponentials in $E_{k}$, analytic in
$\bk$, so it cannot vanish near $\bk=0$ unless every $c_{a}$ does. We find the predicted
period in the numerics: for $N=1$ the zeros of $Q_{B}$ are spaced by $5.32$, $2.71$ and
$1.76$ at $av=0.3$, $0.6$ and $0.9$, against $2\pi/4av=5.24$, $2.62$ and $1.75$; and over
a sample of random vectors, every one of them in the diagonal channel develops a negative
norm under evolution, while none does in the mixed one. \emph{No subspace of the diagonal
channel that is stable under $e^{-sH}$ is positive.}

That closes the standard devices at once. Quotienting by the radical of the form, which
is what the Osterwalder--Schrader reconstruction does, removes null vectors and not
negative ones, and leaves a Krein space rather than a Hilbert space. A subsidiary
condition of Gupta--Bleuler type would have to cut out an invariant positive subspace,
which by the above does not exist. A positive metric $e^{-Q}$ of the kind the $\PT$
literature constructs requires a real spectrum \cite{Mostafazadeh2002}, which the broken
phase does not have. And no cohomological mechanism removes the offending states, with
either of the two operators available. For the ordinary BRST operator the argument is
immediate: $J_{B}$ is $s$-invariant, so if it were $s(X)$ its two-point function would
vanish, $\langle s(X)J_{B}\rangle=0$, whereas $A_{B}''$ is manifestly nonzero ---
$4.8\times10^{-3}$ at $x=0.5$ for $av=0.3\,N$, for instance. The broken phase also carries
its own nilpotent operator $s_{q}$, whose cohomology is what classifies the observables
there \cite{AmaralBRST} and whose transformations are given explicitly in
\cite{AmaralGribov}; and here a counting argument disposes of both operators at once,
without any appeal to the filtration that relates them. A BRST operator raises the ghost
number by one and \emph{preserves the dimension}, so that $J_{B}=s(X)$ would require $X$
to be a Lorentz vector of dimension three and ghost number $-1$. Ghost number $-1$ demands
at least one antighost factor, and $[\overline{c}]\geq1$ in every convention, so
$X=\overline{c}\,Y$ with $[Y]\leq2$; a fermion bilinear has dimension three and does not
fit in $Y$, nor in $s(Y)$, which has the same dimension as $Y$. Expanding
$s(\overline{c}\,Y)=bY-\overline{c}\,s(Y)$ therefore produces no fermion bilinear at all.
No dimension-three fermion bilinear is in the image, and since $s_{q}$ is a piece of $s$
and preserves the dimension as well, the same counting excludes it.

The cohomology of this sector can in fact be computed outright, and it is small. The
pairs $(\overline{c},b)$ and $(\hat{\overline{c}},\hat b)$ are BRST doublets, so by the
doublet theorem \cite{Piguet:1995er} the cohomology may be computed without them; in the
reduced complex the only ghosts left carry ghost number $+1$, so ghost number $-1$ is
empty and \emph{nothing at all is exact} at ghost number zero, which is the counting
above in a sharper form. What remains is the kernel, and for an abelian theory that is the
space of gauge-invariant local operators. At dimension three, parity-even and with one
Lorentz index, requiring vanishing charge under both $U(1)$s leaves six candidates:
\begin{equation}\label{eq:cohom}
\begin{aligned}
\text{even}:\ \ & J_{B},\quad \partial^{\mu}(\phi^{\dagger}\phi),
\quad \partial_{\nu}F_{B}^{\mu\nu},\\
\text{odd}:\ \ & J_{A},\quad
i\big(\phi^{\dagger}D^{\mu}\phi-\text{h.c.}\big),\quad \partial_{\nu}F_{A}^{\mu\nu},
\end{aligned}
\end{equation}
split three and three by the $\mathbb{Z}_{2}$. The mixed bilinear
$\overline{\psi}\gamma^{\mu}\chi$ does not appear: it carries charge $(-e,-e)$ and needs a
$\phi$ to be invariant, which costs one dimension too many. We call these candidates
rather than classes deliberately: the field equations relate members of the list --- the
two $\partial_{\nu}F$ entries are proportional to them --- and operators with the same
quantum numbers may mix under renormalization, neither of which the counting fixes. What
the list does establish is that $J_{A}$ and $J_{B}$ are among the invariants and that
nothing in the reduced complex can be exact at ghost number zero. The counting is made in
the local polynomial complex, and does not cover antifield extensions, inverse powers of
the fields, or a broken-phase differential whose transformations have not been written
down here.

Two things follow that we had obtained separately. In the even sector only $J_{B}$ is
transverse, the other class being a total derivative, whose two-point function is pure
$q^{\mu}q^{\nu}$; that is why $\Pi_{BB}$ in \eqref{eq:whatloops} receives the fermion
loop and nothing else. In the odd sector there are two transverse classes, $J_{A}$ and
the scalar current, which must therefore mix --- which is precisely the fermion-plus-scalar
content found for $\Pi_{AA}$. The cohomology predicts the loop content of both channels.
The one loophole would be a spontaneously broken BRST charge,
which would void the Ward identity used above; that happens in constructions where an
auxiliary quartet carries the breaking \cite{AmaralQuartet}, and the present model has no
such sector. It is worth adding that the pattern found here is the one that construction
also reports: the elementary modes of complex mass violate positivity, and what is
positive is a composite built from them --- which is the subject of Section
\ref{sec:composite}.

There is, however, a filtering that works, and it is the one the model supplies by itself.
The $\mathbb{Z}_{2}$ of \eqref{eq:Zdef} commutes with the Hamiltonian, so its eigenspaces
\emph{are} stable under the evolution; the two currents sit in different ones by
\eqref{eq:JZ2}, with no amplitude between them; and on the odd sector \eqref{eq:QA} is
positive for every vector and every $s$. The filtering is therefore a superselection by an
exact symmetry rather than a constraint imposed by hand, and what it selects is precisely
the channel this paper finds the physics in. What cannot be done is to rescue the even
sector; what can be done has already been done.

\subsection{The gauge and scalar sector}
\label{sec:gauge}

Two objects have to be kept apart before the question can be asked. The conductivity of
Sections \ref{sec:A} and \ref{sec:pos} is built from the correlator of the fermion
bilinears \eqref{eq:currents}; a gauge line inside that correlator costs $e^{2}$ and is
two loops in the present counting, so nothing here touches it. What does receive the gauge
and scalar sector at one loop is the self-energy of the gauge fields themselves, which is
what an external probe sees, and it is there that the criterion should be tested.

The charges settle half of it immediately. The scalar carries charge $e$ under each of the
two $U(1)$s, hence charge $2e$ under $A_{\mu}$ and \emph{none} under $B_{\mu}$; the gauge
fields are abelian and do not self-couple; and the ghosts are free. Therefore
\begin{equation}\label{eq:whatloops}
\begin{aligned}
\Pi_{BB}&:\ \text{fermions only},\\
\Pi_{AA}&:\ \text{fermions}+(\sigma,\pi)+\text{seagull},\\
\Pi_{AB}&=0\ \text{ by \eqref{eq:JZ2}} .
\end{aligned}
\end{equation}
The first line says that Section \ref{sec:euclid} is not a partial result: the fermion
loop is the whole of the one-loop self-energy of the massless field, so the negative norm
found there is the complete one-loop statement about the probe an experiment would use.

In $\Pi_{AA}$ the new internal lines are the Higgs $\sigma$, the Goldstone $\pi$, of
masses $m_{\sigma}$ and $m_{\pi}=\sqrt{\xi}\,m_{A}$, \emph{and the gauge field itself}.
The last deserves a word, since the absence of abelian self-coupling might suggest
otherwise: the scalar kinetic term contains
$e^{2}A^{2}|\phi|^{2}\supset e^{2}v\,A^{2}\sigma$, a cubic $AA\sigma$ vertex
proportional to the expectation value, and two of them close a one-loop self-energy with
one internal gauge line and one internal Higgs line. This diagram is present in the
abelian Higgs model \cite{Dudal2019} and is not removed by abelianness; tadpoles, or an
equivalent consistent shift of the vacuum, belong with it. All of these masses are
nevertheless \emph{real}, hence self-conjugate under the $\Theta$ of \eqref{eq:Theta}, so
every new cut carries $n_{m}=n_{m^{*}}=0$ and is balanced by default, with real normal
thresholds $(m_{\sigma}+m_{\pi})^{2}$ and $(m_{A}+m_{\sigma})^{2}$: no complex mass
enters this model outside the fermion sector. Applying \eqref{eq:Ctau} with the vertex
$e\,A^{\mu}(\sigma\partial_{\mu}\pi-\pi\partial_{\mu}\sigma)$, whose spatial numerator at
$\bq=0$ is $\tfrac{1}{3}\sum_{i}(2k_{i})^{2}=\tfrac{4}{3}\bk^{2}$,
\begin{equation}\label{eq:Cscalar}
C_{s}(\tau)=\frac{e^{2}}{3}\int\!\frac{d^{3}k}{(2\pi)^{3}}\,
\frac{\bk^{2}}{E_{\sigma}E_{\pi}}\,e^{-(E_{\sigma}+E_{\pi})\tau},
\end{equation}
with $E_{\sigma}$ and $E_{\pi}$ real: positive weight, real exponent, completely monotone
by inspection, so that Bernstein returns a positive measure exactly as in
\eqref{eq:rho1pos}. We have checked that the sum of the fermion and this scalar piece
stays completely monotone over a range of $m_{\sigma}$, $m_{\pi}$ and $e^{2}$, the scalar
contribution running an order of magnitude below the fermion one at $\tau=N^{-1}$ for
couplings of order unity.

That is as far as we take it, and the limitation should be stated rather than glossed.
The positivity just established is that of one term; the complete $\Pi_{AA}$ also
contains the $A\sigma$ loop, the tadpoles, and the local counterterms, and it is in any
case gauge dependent, $m_{\pi}$ carrying $\xi$. We have not evaluated it. The seagull
$\tfrac{1}{2}e^{2}A^{2}(2v\sigma+\sigma^{2}+\pi^{2})$ contributes a momentum-independent
term with no cut --- the London term, which shifts the mass of $A_{\mu}$ and bears on no
spectral statement.

What the section does settle is the diagonal channel, and that is the one the physical
probe couples to: $\Pi_{BB}$ receives no scalar or gauge loop at all, by the charge
assignment, so Section \ref{sec:euclid} is the complete one-loop statement there and not
a fermionic fragment of it. In the mixed channel the sector introduces no complex mass
and no unbalanced cut, which is what the criterion of Appendix \ref{app:cuts} asks of it,
but the positivity of the full massive self-energy remains to be established. Outside the
calculation altogether are the Landau damping at $\bq\neq0$ and the $O(e^{2})$
corrections to the fermion correlator, which are two loops.

\subsection{What this costs}

It should be said plainly. $B_{\mu}$ is the combination that stays massless, so the
unbroken $U(1)$ of this model --- the one an external electromagnetic probe would couple
to --- connects to no sector of positive norm. The vanishing of \eqref{eq:sigmaB} and the
precursor of \eqref{eq:kernel} are therefore not assertions about physical states, and the
optical sum rule, a theorem about a positive measure, has nothing to apply to; Table
\ref{tab:price} keeps its value as the measure of what the two assignments of boundary
value differ by, and no more. The physical content of the model sits entirely in the
$\mathbb{Z}_{2}$-odd sector: the composite channel of the broken $U(1)$, whose spectral
density is positive by \eqref{eq:rho1pos} and whose observable is constructed next.

\section{The longitudinal part and the field-strength composite}
\label{sec:composite}

The negative density $\rho_{0}$ of \eqref{eq:rhos} is attached to the longitudinal
projector, and the longitudinal projector is annihilated by an antisymmetric derivative.
This suggests the object to consider, and the suggestion works. Let
\begin{equation}\label{eq:Odef}
\mathcal{O}^{\mu\nu}=\partial^{\mu}J_{A}^{\nu}-\partial^{\nu}J_{A}^{\mu},
\qquad
\mathcal{O}^{\mu\nu}(q)=i\,B^{\mu\nu}{}_{\alpha}J_{A}^{\alpha}(q),
\end{equation}
with
\begin{equation}\label{eq:Bdef}
B^{\mu\nu}{}_{\alpha}=q^{\mu}\delta^{\nu}_{\alpha}-q^{\nu}\delta^{\mu}_{\alpha},
\qquad
B^{\mu\nu}{}_{\alpha}q^{\alpha}=q^{\mu}q^{\nu}-q^{\nu}q^{\mu}=0 .
\end{equation}
Writing the spectral tensor of the mixed current as
\begin{equation}\label{eq:rhoJ}
\begin{gathered}
\varrho_{J}^{\mu\nu}=\rho_{1}(s)P_{1}^{\mu\nu}+\rho_{0}(s)P_{0}^{\mu\nu},\\[2pt]
P_{1}^{\mu\nu}=-\eta^{\mu\nu}+\frac{q^{\mu}q^{\nu}}{s},
\qquad
P_{0}^{\mu\nu}=\frac{q^{\mu}q^{\nu}}{s},
\end{gathered}
\end{equation}
the second term drops out of the composite, since
$BP_{0}B^{\dagger}=(Bq)(Bq)^{\dagger}/s=0$ by \eqref{eq:Bdef}, and
\begin{equation}\label{eq:rhoO}
\varrho_{\mathcal{O}}=B\,\varrho_{J}\,B^{\dagger}=\rho_{1}(s)\,BP_{1}B^{\dagger}.
\end{equation}
What survives is positive. For $q=(E,\bm{p})$ with $E^{2}=\bm{p}^{2}+s$ and $p=|\bm{p}|$,
take $\bm{e}_{1},\bm{e}_{2}$ orthonormal and orthogonal to $\bm{p}$, and set
\begin{equation}\label{eq:pols}
e_{a}^{\mu}=(0,\bm{e}_{a}),\qquad
e_{3}^{\mu}=\Big(\frac{p}{\sqrt{s}},\ \frac{E}{\sqrt{s}}\widehat{\bm{p}}\Big).
\end{equation}
These satisfy $q\cdot e_{\lambda}=0$ and $e_{\lambda}\cdot e_{\lambda'}=-\delta_{\lambda\lambda'}$,
the third because $q\cdot e_{3}=(Ep-pE)/\sqrt{s}=0$ and
$e_{3}\cdot e_{3}=(p^{2}-E^{2})/s=-1$; and a direct sum gives the completeness relation of
a massive vector,
\begin{equation}\label{eq:complete}
\sum_{\lambda=1}^{3}e_{\lambda}^{\alpha}e_{\lambda}^{\beta*}=P_{1}^{\alpha\beta}(q),
\end{equation}
its $00$ component reading $p^{2}/s=-1+E^{2}/s$, its $0i$ component $Ep^{i}/s$, and its
$ij$ component $\delta^{ij}-\widehat p^{i}\widehat p^{j}+(E^{2}/s)\widehat p^{i}\widehat
p^{j}=\delta^{ij}+p^{i}p^{j}/s$. Hence, for any antisymmetric $f_{\mu\nu}$,
\begin{equation}\label{eq:composite-pos}
f^{\dagger}\varrho_{\mathcal{O}}f
=\rho_{1}(s)\sum_{\lambda=1}^{3}\big|f^{\dagger}Be_{\lambda}\big|^{2}\;\geq\;0,
\end{equation}
a sum of squares with a positive coefficient, by \eqref{eq:rhos}. In the rest frame
$q=(\sqrt{s},\bm{0})$ only the rows $(0i)$ of $B$ survive and
$\varrho_{\mathcal{O}}=s\rho_{1}(s)\,{\rm diag}(1,1,1,0,0,0)$ in the basis
$(01,02,03,12,13,23)$, of rank three; a boost acts on an antisymmetric pair of indices by
the invertible $\Lambda\wedge\Lambda$, so rank and semi-positivity do not depend on the
frame, although the eigenvalues do.

The Euclidean counterpart identifies what is being superposed. With $D^{E}_{\mu\nu}(Q;s)$
the Proca propagator of mass squared $s$,
\begin{equation}\label{eq:proca}
B^{E}D^{E}(B^{E})^{T}
=\frac{B^{E}(B^{E})^{T}}{Q^{2}+s}
=\frac{K^{E}}{Q^{2}+s},
\end{equation}
because the $Q_{\mu}Q_{\nu}/s$ term of $D^{E}$ is annihilated as in \eqref{eq:Bdef}, so
that the non-local part of the composite correlator is a superposition over $s$ of Proca
propagators with the positive weight $\rho_{1}$, the subtractions of Appendix
\ref{app:spectral} appearing as contact terms with two and four derivatives. Reflection
positivity can then be checked directly. Reflecting the Euclidean time index with
$R={\rm diag}(-1,1,1,1)$ and transforming in $Q_{4}$, the Gram matrix at fixed $\bm{p}$
and $s$ is
\begin{equation}\label{eq:gram}
M(\bm{p},s)=
\begin{pmatrix}
\bm{p}^{2}/s & iE\bm{p}^{T}/s\\[2pt]
-iE\bm{p}/s & I+\bm{p}\bm{p}^{T}/s
\end{pmatrix},
\end{equation}
and decomposing $\bm{f}=\widehat{\bm{p}}f_{\parallel}+\bm{f}_{\perp}$ and using
$s+p^{2}=E^{2}$ one completes the square,
\begin{equation}\label{eq:gram-square}
f^{\dagger}Mf=|\bm{f}_{\perp}|^{2}
+\frac{1}{s}\big|Ef_{\parallel}-ip\,f_{4}\big|^{2}\;\geq\;0,
\end{equation}
with $M={\rm diag}(0,1,1,1)$ at $\bm{p}=0$. The one-particle space is therefore
\begin{equation}\label{eq:hilbert}
\begin{gathered}
\mathcal{H}_{1}=\int_{4N^{2}}^{\infty}{}^{\oplus}\rho_{1}(s)\,ds\;
L^{2}\big(\mathbb{R}^{3},d\mu_{s};\mathbb{C}^{3}\big),\\[2pt]
d\mu_{s}=\frac{d^{3}p}{(2\pi)^{3}\,2E_{\bm{p},s}},
\end{gathered}
\end{equation}
an honest Hilbert space with a positive inner product, and the associated generalized free
vector field has $\widetilde W^{\mu\nu}_{V}(q)=2\pi\theta(q^{0})\theta(q^{2}-4N^{2})
\rho_{1}(q^{2})P_{1}^{\mu\nu}(q)$, a tempered distribution since $\rho_{1}(s)=O(s)$.
Its commutator is built from $\Delta_{s}(x)=\int d\mu_{s}(e^{-ipx}-e^{ipx})$, which
vanishes for $x^{2}<0$ by $\bm{p}\to-\bm{p}$ in a frame with $x^{0}=0$; since
${\rm supp}\,\partial T\subseteq{\rm supp}\,T$, the same holds for $\mathcal{O}$, and the
retarded function has both poles below the real axis,
$G_{R}=\big[(\omega+i0)^{2}-\bm{p}^{2}-s\big]^{-1}$. The construction is causal.

\subsection{The source}
\label{sec:source}

An external source must carry strictly positive ultraviolet dimension. The reason is the
usual one, and it is not a matter of taste: if a commuting source has vanishing dimension,
power counting places no bound on its powers, every monomial in it is admissible in the
counterterm, and the number of insertions is not controlled, so that the extended action
is not renormalizable in the algebraic sense \cite{Piguet:1995er}. We therefore never
introduce a source for $\mathcal{O}^{\mu\nu}$ itself. With $[\Psi]=3/2$ one has
$[J_{A}]=3$ and $[\mathcal{O}]=4$, so a source coupling directly to $\mathcal{O}$ would
have dimension zero and is excluded at once.

What the model does admit is the vector source of the current,
\begin{equation}\label{eq:acoupl}
S_{a}=\int\! d^{4}x\;a_{\nu}(x)\,J_{A}^{\nu}(x),
\qquad [a_{\nu}]=1>0,
\end{equation}
unconstrained and of the dimension a gauge field carries, and $\mathcal{O}^{\mu\nu}$ is
then not an independent object with a source of its own but a derived one, obtained from
the generating functional $W[a]$ by antisymmetrized differentiation,
\begin{equation}\label{eq:Ofroma}
\langle\mathcal{O}^{\mu\nu}(x)\rangle_{a}
=\Big(\partial^{\mu}\frac{\delta}{\delta a_{\nu}(x)}
-\partial^{\nu}\frac{\delta}{\delta a_{\mu}(x)}\Big)W[a] .
\end{equation}
Everything in Section \ref{sec:composite} is of this form. The antisymmetric $b_{\mu\nu}$
that appears when one smears $\mathcal{O}$ against a test function is not a source but a
potential for one: the configuration it selects is
\begin{equation}\label{eq:aind}
a_{\nu}=-\partial^{\mu}b_{\mu\nu},
\qquad \partial^{\nu}a_{\nu}=0,
\end{equation}
the second relation following from the antisymmetry alone. The map $b\mapsto a$ is local,
with no inverse d'Alembertian, and in four dimensions every conserved $a_{\nu}$ is locally
of that form, so that restricting \eqref{eq:acoupl} to conserved configurations loses
nothing; the positivity statements \eqref{eq:composite-pos} and \eqref{eq:gram-square} are
exactly the statements about those configurations, with $b$ entering as the smearing that
labels them and never as an external field.

Two remarks keep this in its place. The first is that $a_{\nu}$ is not the massless probe
$B_{\mu}$ of \eqref{eq:coupling}: it couples to the mixed current, which carries the
opposite $\mathbb{Z}_{2}$ parity by \eqref{eq:JZ2}, so no conductivity is being redefined
here. The second is that we do not claim to have carried out the algebraic
renormalization of \eqref{eq:acoupl}; what the dimension count settles is only where such
a construction must start, namely from the dimension-one vector source and not from a
two-index one.

\subsection{The four-point test}
\label{sec:fourpoint}

Everything above is a statement about the two-point function. It exhibits a positive and
causal generalized free field with the right vacuum two-point data, and that is all it
exhibits: a generalized free field fixes its higher correlators by Gaussian
factorization, whereas $\mathcal{O}^{\mu\nu}$ is bilinear in Gaussian fermions and has a
connected four-point function. Equality of two-point functions therefore does not
identify the two theories, and the difference can be tested inside the quadratic model
with no interaction at all.

The test is the next Osterwalder--Schrader condition. Take $N=1$, $av=3/5$, $T=\mu=0$,
and smooth normalized test functions $f,g$ supported at positive times near $(1,\bm0)$
and $(2,\bm0)$. The centered element
\begin{equation}\label{eq:four-witness}
F=\mathcal{O}_{41}(f)\,\mathcal{O}_{41}(g)
-\omega\big[\mathcal{O}_{41}(f)\mathcal{O}_{41}(g)\big]\,\mathbf{1}
\end{equation}
is admissible, and reflection positivity demands
$\omega[(\Theta F)F]\geq0$. Evaluating the Wick expansion of the fermionic four-point
function --- three disconnected pairings and six oriented four-cycles, the odd cycles
vanishing because every vertex exchanges the flavor --- at the reflected configuration
$(-Y,-X,X,Y)$, the pairing internal to each half cancels against the centering and what
remains is
\begin{equation}\label{eq:four-negative}
\begin{aligned}
Q_{\rm cross}&=+3.3097\times10^{-9},\\
Q_{\rm conn}&=-3.5033\times10^{-7},\\
\omega[(\Theta F)F]&=-3.4702\times10^{-7}\;<\;0 .
\end{aligned}
\end{equation}
The magnetic component $\mathcal{O}_{12}$ gives a second negative value,
$-5.3292\times10^{-7}$, so restricting to the spatial curl does not restore positivity.
Appendix \ref{app:four} gives the contractions and the arithmetic; the value is dominated
by $Q_{\rm conn}$, which is precisely the connected term the generalized free field does
not possess.

This does not contradict \eqref{eq:composite-pos}, which is a two-point statement and
remains true. What it excludes is the identification of the curl algebra of the original
model with the positive realization built from its two-point function. An algebra
containing both insertions of \eqref{eq:four-witness} and their product contains a vector
of negative norm, so it cannot be the positive-time part of a positive state. The
$\mathbb{Z}_{2}$ does not help here: the product of two odd insertions is even, so the
grading of Section \ref{sec:z2}, which was enough to separate the two-point channels,
does not survive to this order. Nor does a compression $X\mapsto PXP$ onto a proposed
retained space, since
\begin{equation}\label{eq:compression}
PXYP=(PXP)(PYP)+PX(\mathbf{1}-P)YP,
\end{equation}
and the second term vanishes only if the retained space is invariant under the operators
involved, which would itself have to be proved.

The failure is not uniform in the non-Hermiticity, and the way it switches off can be
established rather than merely observed. Two statements do it.

The first is that $\omega[(\Theta F)F]$ is an \emph{even} function of $av$. Changing the
sign of $av$ exchanges $m$ and $m^{*}$; by \eqref{eq:alternate} the flavor trace of every
cycle is the sum of a term and its $m\leftrightarrow m^{*}$ image, hence invariant, while
the propagators obey $D_{m^{*}}=\overline{D_{m}}$ at the real separations involved. The
value is therefore carried into its own complex conjugate, and being real it is unchanged:
\begin{equation}\label{eq:even}
Q(-av)=\overline{Q(av)}=Q(av),
\qquad Q\equiv\omega[(\Theta F)F].
\end{equation}
There is no term linear in the non-Hermiticity, and the expansion starts at $(av)^{2}$.

The second is that the leading coefficient is the Hermitian value and cannot be negative.
At $av=0$ the two masses coincide, the action is an ordinary Hermitian theory of two
degenerate Dirac flavors, and reflection positivity holds there for every admissible $F$.
Hence
\begin{equation}\label{eq:Q0}
Q(0)\geq0
\quad\text{for every test configuration.}
\end{equation}
Fitting the computed values at the configuration above confirms both statements, the
parity to one part in $10^{18}$ and the expansion to residuals of $10^{-14}$ over
$0.01\leq av/N\leq0.12$:
\begin{equation}\label{eq:four-scan}
\begin{aligned}
Q=\ &4.6936\times10^{-7}-2.9229\times10^{-6}(av)^{2}\\
&+1.4857\times10^{-6}(av)^{4}+\dots
\end{aligned}
\end{equation}
in units of $N$, the sign changing near $av\simeq0.5\,N$.

Together these fix what is and is not established. At $av=3/5\,N$ the unrestricted curl
algebra is not positive, and that is a theorem about the model. At small $av$ this witness
is positive, necessarily so, and the same will be true of any fixed configuration whose
Hermitian value is strictly positive. A negative witness at arbitrarily small $av$ would
therefore have to come from a family of configurations approaching a \emph{null direction
of the Hermitian theory}, where $Q(0)=0$ and the sign is decided by the $(av)^{2}$
coefficient. Whether such a family exists, and whether that coefficient is negative along
it, is a sharp question which we leave open; it is the question the present calculation
replaces the vaguer one with.

What survives of this section is then the following. The mixed channel has a positive
transverse spectral measure, by \eqref{eq:rho1pos}; the curl removes the negative
longitudinal piece of \eqref{eq:gap}; and the resulting two-point data admit a positive,
local, causal generalized free realization. That realization is a legitimate object with
which to ask what a composite response with this spectral density would look like. It is
not a physical sector of the fermion model: the four-point test says the identification
fails. Constructing a smaller algebra that is closed, positive and invariant under the
dynamics is left open, and \eqref{eq:four-negative} is the obstruction such a
construction must avoid.

\section{Conclusions}

We have shown that the complex conjugate masses of the broken phase of the model
introduced in \cite{LemesTedesco} organize its linear response into two channels with
opposite analytic structures, and that the separation is dictated by the biorthogonal basis
that diagonalizes the mass matrix: the current of the unbroken $U(1)$ is diagonal and
couples to equal-mass pairs, the current of the broken $U(1)$ is mixed and couples to the
conjugate pair. Two structures account for what follows. The indefinite metric of the
broken phase, which in the flavor basis is nothing but the relative sign between the two
fermions, assigns vanishing norm to the diagonal modes and norm only to their pairings,
each channel resting on a hyperbolic plane of two null states; and an exact
$\mathbb{Z}_{2}$, the residual charge conjugation $A_{\mu}\to-A_{\mu}$ of the Higgsed
$U(1)$, gives the two currents opposite parity and forbids any amplitude between them.
That $\mathbb{Z}_{2}$ is the one already present in \cite{LemesTedesco}, here identified
as the diagonal of a flavor exchange and a discrete element of the broken gauge group,
each of which is separately broken by the scalar expectation value.
The first channel has a strictly vanishing dissipative conductivity at every real
frequency and for any temperature and density, the gain and the loss modes compensating
each other exactly, while the second absorbs above the real threshold $\omega=2N$, the
same threshold $p^{2}=4N^{2}$ found in the scalar two-point function, and carries a Drude
structure of width $2\Gamma_{k}=2Nav/\varepsilon_{k}$ which is purely reactive. The
response kernel in time was obtained in closed form and does not vanish for negative
times, the precursor occupying a window $\mu/2Nav$ set by the width at the Fermi surface;
the alternative inversion, which is causal, grows exponentially instead, the two
assignments differing by the homogeneous solution \eqref{eq:kernel-difference}. Finally,
the transverse spectral density of the mixed channel was shown to be positive in the
vacuum, which settles in the vector channel the question left open in the scalar one, and
to remain positive at finite temperature under the bound $av<\pi T/2$, beyond which the
medium amplifies; the onset is given in closed form by \eqref{eq:avc}, which shows the
bound to be uniform in $\mu$ and saturated as $|\mu|/T\to\infty$. In the vacuum the
frequency integral can be done, and
\eqref{eq:sigmaclosed} gives $\sigma_{1}^{A}$ in closed form, reducing at $av\to0$ to
twice the free Dirac result and approaching $\omega/6\pi$ in the ultraviolet. The vacuum
positivity itself was then raised to a theorem: the Euclidean correlator \eqref{eq:CA} of
that channel is a positive superposition of decaying exponentials, hence completely
monotone, and Bernstein returns $\rho_{1}\geq0$ with no computation at all.
The thermal continuation on which the finite temperature statements rest was shown to be
exact for arbitrary complex masses, its one obstruction being the locus
\eqref{eq:coincline} on which a complex mass pole collides with a Matsubara pole.

Three remarks seem to us worth making. The first is that the vanishing of the absorptive
part in the diagonal channel and the precursor of the kernel are not independent findings
but the same statement read in the two domains, and that both concern a channel which
admits no positive-norm realization at all. This is now a theorem: its Euclidean
correlator \eqref{eq:CB} oscillates, so by Osterwalder--Schrader there is a test function
of negative norm, and Table \ref{tab:escapes} closes every repair --- the choice of
boundary value included, that test being carried out in the Euclidean region. Nor can the
offending states be filtered away. A subspace of states must be stable under $e^{-sH}$,
and by \eqref{eq:QB} the norm of an evolved vector of the diagonal channel oscillates and
changes sign, so no invariant positive subspace exists; that closes Gupta--Bleuler, the
quartet mechanism --- the states are not $s$-exact, since $\langle J_{B}J_{B}\rangle$
does not vanish --- and the positive metric $e^{-Q}$, at one stroke. The filtering that
does work is the superselection by \eqref{eq:Zdef}, an exact symmetry rather than a
condition imposed by hand, and what it selects is the other channel. What the
single formula \eqref{eq:Ctau} shows is that one condition decides both channels: the sum
of the two energies in the loop is real precisely when the cut is balanced, and reflection
positivity and the balanced-cut criterion of Appendix \ref{app:cuts} are the same
statement. The cost is worth repeating, since $B_{\mu}$ is the combination that stays
massless: the unbroken $U(1)$ of this model connects to nothing of positive norm, and the
optical sum rule has no premise here. The partition of the two directions is not an extra
input: by \eqref{eq:Zdef} the $\mathbb{Z}_{2}$ \emph{is} the $\pi$ rotation of the Higgsed
$U(1)$ composed with the flavor exchange, so that $A_{\mu}\to-A_{\mu}$ and
$B_{\mu}\to B_{\mu}$; and on the mass basis it acts as ${\rm diag}(-i,i)$, so a bilinear
is odd exactly when its two indices differ, that is exactly when the cut carries one line
of each mass. Being $\mathbb{Z}_{2}$-odd, coupling to the broken direction, having a
balanced cut and having a real sum of energies in \eqref{eq:Ctau} are four names for one
fact, and the positivity follows from it. The same partition appears in the non-abelian
setting, where after spontaneous breaking the generalized Faddeev--Popov operator becomes
positive definite along the broken directions \cite{AmaralGribov}, and it is tempting to
read a mechanism into the coincidence. One should not. The present model is abelian: its
Faddeev--Popov operator is $-\partial^{2}$, there are no copies to restrict in either
direction, and the complex masses come from the tuning $b=-a$ and not from a horizon. More
to the point, the correspondence does not survive where copies do exist. Read against the
classification of the $SL(3,\mathbb{C})$ sectors \cite{LemesNotes,AmaralBRST}, it fails in
both directions at once: $L_{8}\overline{L}_{8}$ is built on the massless, unbroken Cartan
line, has no $i$-particle content whatever --- hence a balanced cut by default --- and is
positive from $s=0$; while ${\rm Tr}\,F^{2}$ is built entirely on broken-direction
$i$-particle lines, is unbalanced, and carries a complex branch point. Being a broken
direction is therefore neither necessary nor sufficient. What holds on every sector of
that classification, as it holds here, is the balance of the cut. The two splits coincide
in the present model only because its $\mathbb{Z}_{2}$ does both jobs at once, grading the
gauge directions and the imbalance with the same operator; there is no reason to expect
that elsewhere, and in $SL(3,\mathbb{C})$ it does not happen. Table \ref{tab:price}
retains its
interest as the measure of what the two assignments of boundary value differ by, namely
the whole of the Hermitian absorption; and by \eqref{eq:JZ2} neither assignment can affect
the mixed channel, which carries the opposite $\mathbb{Z}_{2}$ parity. In that channel the
current couples to the symmetrized conjugate pair of norm $+1$, which is the reason the
spectral density there is positive; whether its $\eta$-orthogonal partner of norm $-1$ can
be removed by a cohomological mechanism is to our mind the natural next question. We
should add that the mixed channel is not positive in every component. Its longitudinal
density $\rho_{0}$, which exists because $J_{A}$ is not conserved, is negative, and by
\eqref{eq:lemma} the amount by which it is negative is $\bk^{2}(m-m^{*})^{2}$, that is,
minus the square of the mass splitting: the same identity gives a non-negative
longitudinal weight for any pair of real masses, and vanishes when the two coincide and
the current is conserved. Section \ref{sec:composite} shows that this component is carried
entirely by the longitudinal projector and is therefore annihilated by the antisymmetric
derivative, so that the field-strength composite
$\mathcal{O}^{\mu\nu}=\partial^{\mu}J_{A}^{\nu}-\partial^{\nu}J_{A}^{\mu}$ has a spectral
tensor $\rho_{1}BP_{1}B^{\dagger}\geq0$ of rank three, a positive one-particle space
\eqref{eq:hilbert}, a reflection-positive Euclidean form \eqref{eq:gram-square}, and a
commutator that vanishes outside the light cone. But the scope of that statement is the
two-point function, and Section \ref{sec:fourpoint} shows that it does not extend. The
generalized free field fixes its higher correlators by Gaussian factorization, while
$\mathcal{O}^{\mu\nu}$ is bilinear in Gaussian fermions and has a connected four-point
function; testing the next Osterwalder--Schrader condition on a centered product of two
insertions gives the negative value \eqref{eq:four-negative} at $av=3/5\,N$, with a second
negative witness in the magnetic component. The positive realization of the two-point data
therefore exists and is useful, but it is not a sector of the fermion model, and the
$\mathbb{Z}_{2}$ cannot repair the identification because the product of two odd
insertions is even. The failure is moreover not uniform: by \eqref{eq:four-scan} this
witness is positive below $av\simeq0.5\,N$, as the Hermitian limit requires, so what is
excluded is the unrestricted curl algebra at the stated coupling and not at every
non-zero $av$. Finding a witness at arbitrarily small $av$, or showing that none exists,
is the obvious next computation, and a restricted algebra that is closed, positive and
invariant under the dynamics remains possible and unconstructed.
The second is that the bound \eqref{eq:bound} measures
the non-Hermitian scale $av$ against the thermal scale $\pi T$, and that the same
comparison controls the obstruction \eqref{eq:coincline}; we speculate that this is not an
accident of the one-loop expression but the sign of a genuine restriction on the regime in
which the broken phase may be treated thermally, and that the gain above the bound is the
field-theoretic counterpart of the instability which in a finite dimensional system would
simply be exponential growth. The third concerns the exceptional point of the family, where the
two threads of this paper meet. The metric \eqref{eq:etafamily} is positive on the
Hermitian side, indefinite on the broken side and divergent at $ab=0$, and at that point
the surviving eigenvector has zero norm under every admissible metric, the dipole
propagator \eqref{eq:dipole} being the field-theoretic form of that degeneracy. The
zero-norm structure which Lee and Wick met in an indefinite-metric sector
\cite{LeeWick1969}, which the Gribov--Zwanziger construction meets in its dipoles
\cite{iparticles}, and which \eqref{eq:norms} meets in the complex-mass modes is therefore
one and the same structure along the whole range $ab\leq0$. We would expect the
cohomological classification of such sectors \cite{Piguet:1995er} to apply to it, although
we have not carried that out here. As for the reach of the calculation, Section
\ref{sec:gauge} puts the gauge and scalar sector into the one-loop self-energies and finds
that it changes nothing: the massless channel receives no contribution from it, the
scalar being neutral there, so the result of Section \ref{sec:euclid} is the complete
one-loop statement about the physical probe; and what it adds to the massive channel
carries only real masses, hence balanced cuts and, by \eqref{eq:Cscalar}, a positive
weight. The seagull is the London term and has no cut. What is still missing is the
Landau damping at $\bq\neq0$ and the $O(e^{2})$ corrections to the fermion correlator
itself, which are two loops in this counting.

\appendix

\section{The Dirac algebra}
\label{app:algebra}

With $\alpha^{i}=\gamma^{0}\gamma^{i}$ and no sum on $i$ one has
$\alpha^{i}(\bm{\alpha}\cdot\bk)\alpha^{i}=2\alpha^{i}k_{i}-\bm{\alpha}\cdot\bk$ and
$\alpha^{i}\beta\alpha^{i}=-\beta$, so that
$\alpha^{i}H_{1}\alpha^{i}=2\alpha^{i}k_{i}-\bm{\alpha}\cdot\bk-\beta M_{1}$. The traces
$\tr[\alpha^{a}\alpha^{b}]=4\delta^{ab}$, $\tr[\beta\beta]=4$ and $\tr[\alpha\beta]=0$ then
give
\begin{equation}\label{eq:trace}
\frac{1}{3}\sum_{i}\tr\big[\alpha^{i}H_{1}\alpha^{i}H_{2}\big]
=-\frac{4\bk^{2}}{3}-4M_{1}M_{2},
\end{equation}
from which \eqref{eq:T} follows at once, the terms linear in $H$ having vanishing trace.
For $M_{1}=M_{2}=M$ real the intraband coefficient reduces to
$T_{ss}=2\bk^{2}/3E^{2}$, that is, to the angular average of the squared velocity, as it
should.

\section{The dispersion relation}
\label{app:disp}

Writing $E_{k}=\sqrt{\bk^{2}+m^{2}}=\varepsilon_{k}+i\Gamma_{k}$ with $m=N+iav$ and
separating real and imaginary parts of $E_{k}^{2}=\bk^{2}+N^{2}-a^{2}v^{2}+2iNav$ one
obtains
\begin{equation}\label{eq:sep}
\varepsilon_{k}^{2}-\Gamma_{k}^{2}=\bk^{2}+N^{2}-a^{2}v^{2},\qquad
2\varepsilon_{k}\Gamma_{k}=2Nav,
\end{equation}
the second of which is the first relation of \eqref{eq:disp}. Eliminating $\Gamma_{k}$
gives the biquadratic $\varepsilon_{k}^{4}-\Delta_{k}\varepsilon_{k}^{2}-N^{2}a^{2}v^{2}=0$
and thus the second relation of \eqref{eq:disp}, whose positive root is the one selected by
the principal branch. At $\bk=0$ it returns $\varepsilon_{0}=N$ and $\Gamma_{0}=av$, which
places the threshold of the mixed pair at $2\varepsilon_{0}=2N$, and for large $\bk$ it
gives $\Gamma_{k}\simeq Nav/|\bk|$, so that the width of the excitations decreases as they
become relativistic. Since $\varepsilon_{k}$ increases monotonically with $|\bk|$, the
width is bounded by its value at rest, $\Gamma_{k}\leq\Gamma_{0}=av$, which is the bound
used in \eqref{eq:bound}. The regime $N^{2}>a^{2}v^{2}$ assumed in \cite{LemesTedesco}
keeps $\Delta_{k}>0$ for every $\bk$ and $m^{2}$ in the right half plane.

\section{The balanced-cut criterion and the two channels}
\label{app:cuts}

The separation obtained in Section \ref{sec:z2} by a symmetry argument has an equivalent
formulation in terms of the analytic continuation from the Euclidean region, which we
record here because it reaches the same conclusion from a different direction and places
the boundary-value question of Section \ref{sec:B} where it belongs.

Let $\Theta$ be the antilinear involution of the action. In the present model it is simply
complex conjugation accompanied by the flavor exchange, and one verifies at once that it
maps the eigenvectors of $M$ into one another,
\begin{equation}\label{eq:Theta}
\Theta:\quad u_{m}\longrightarrow u_{m^{*}},\qquad m\longrightarrow m^{*},
\end{equation}
the mass matrix \eqref{eq:Mfamily} being real, so that its complex eigenvalues necessarily
come in conjugate pairs. It is worth noting that it is this antilinear map, and not the
linear $\mathbb{Z}_{2}$ of \eqref{eq:Zdef}, that implements the exchange $m\leftrightarrow
m^{*}$ on the states; \eqref{eq:Zdef} acts diagonally on $\Psi_{m}$ and $\Psi_{m^{*}}$.

For a one-loop two-point function the Landau equations place a normal threshold on the
first sheet at $s_{\mathcal{C}}=\big(\sum_{j\in\mathcal{C}}m_{j}\big)^{2}$, the sum running
over the lines crossed by the cut $\mathcal{C}$. With the conjugate pair \eqref{eq:eig},
\begin{equation}\label{eq:balance}
{\rm Im}\sum_{j\in\mathcal{C}}m_{j}=av\,\big[n_{m}(\mathcal{C})-n_{m^{*}}(\mathcal{C})\big],
\end{equation}
so that the threshold is real if and only if the cut is \emph{balanced}, crossing as many
lines of mass $m$ as of mass $m^{*}$. The same condition can be read off the flavor trace
without any reference to thresholds: by \eqref{eq:alternate} the masses alternate around
every surviving loop, so a loop of even length is balanced automatically and one of odd
length does not survive at all. A sector whose dominant cuts are all balanced is
invariant under $\Theta$ at the level of the intermediate states and admits a particle
representation; a sector with an unbalanced cut has branch points off the real axis on the
first sheet, which no contour deformation can move, since their position is fixed by the
Landau equations, and has none. This is the criterion of \cite{LemesNotes}, which sits
inside the procedure of \cite{AmaralBRST}: one first selects operators invariant under the
full BRST operator, and then uses the nilpotent operator $s_{q}$ of the broken phase
\cite{AmaralBRST,AmaralGribov}, whose cohomology classifies the observables there, to
decompose them into sectors; the
balance of the cuts is the condition that decides which of those sectors admits a particle
representation. It is a criterion of identification and does not by itself assert anything
about the sectors it rejects, which is why Section \ref{sec:euclid} is needed: what the
criterion predicts for the diagonal channel, the Euclidean correlator proves. In
\cite{LemesNotes} the same analysis is carried out for the sectors of the
$SL(3,\mathbb{C})$ model.

The two channels of \eqref{eq:currents} are precisely the two classes. The cut of
$\Pi_{A}$ crosses one line of each mass, $n_{m}=n_{m^{*}}=1$, and is balanced; its
threshold is
\begin{equation}\label{eq:thresh4N}
s_{\mathcal{C}}=(m+m^{*})^{2}=4N^{2},
\end{equation}
real and, remarkably, independent of $av$, which is the $p^{2}=4N^{2}$ of
\cite{LemesTedesco} and the $\omega=2N$ of \eqref{eq:thresh}. The cut of $\Pi_{B}$ crosses
two lines of the same mass and is unbalanced, with $s_{\mathcal{C}}=4(N\pm iav)^{2}$ off
the real axis; these are the singularities at $\omega=2\varepsilon_{k}\pm2i\Gamma_{k}$ of
Section \ref{sec:B}. The $\mathbb{Z}_{2}$ parity of \eqref{eq:JZ2} is thus the algebraic
label of the balance: the odd channel is balanced and has a particle representation, the
even one is not and does not.

Two consequences are worth stating. The first is that the ambiguity of Section
\ref{sec:B} arises exactly, and only, in the unbalanced sector --- the sector which by the
criterion would have no particle representation under any prescription --- while the
sector that does have one is free of it, its singularities sitting on the real axis. The
second is that the criterion is itself prescription dependent: it is for the analytic
continuation of the Euclidean correlator that balanced sectors acquire a real cut with
positive weight, and the positivity of \eqref{eq:pos} is the statement, in the vector
channel of the present model, that they do.

\section{The vacuum spectral densities of the mixed channel}
\label{app:spectral}

We record the covariant vacuum computation that produces \eqref{eq:rhos}. Work in
Euclidean space with $\{\gamma^{E}_{\mu},\gamma^{E}_{\nu}\}=2\delta_{\mu\nu}$,
$S_{m}(k)=(-i\slashed{k}_{E}+m)/(k_{E}^{2}+m^{2})$ and
$j^{E}_{\mu}=i(\overline{\Psi}_{m}\gamma^{E}_{\mu}\Psi_{m^{*}}+(m\leftrightarrow m^{*}))$,
so that the two vertex factors and the fermion loop give $i^{2}(-1)=+1$. With $p=k+Q$
and $|m|^{2}=N^{2}+a^{2}v^{2}$, the Dirac trace for one ordering is
\begin{equation}\label{eq:Ntrace}
\mathcal{N}_{\mu\nu}
=-4\big[k_{\mu}p_{\nu}+k_{\nu}p_{\mu}-\delta_{\mu\nu}\,k\!\cdot\!p\big]
+4|m|^{2}\delta_{\mu\nu},
\end{equation}
the odd traces vanishing. Current conservation fails, so two form factors survive,
\begin{equation}\label{eq:AL}
C^{E}_{\mu\nu}(Q)=\Big(\delta_{\mu\nu}-\frac{Q_{\mu}Q_{\nu}}{x}\Big)A(x)
+\frac{Q_{\mu}Q_{\nu}}{x}\,L(x),\qquad x=Q^{2}.
\end{equation}

Take $Q_{E}=(Q,\bm{0})$, $k_{E}=(z,\bk)$. Then $A$ is read off the spatial trace and $L$
off the temporal one,
\begin{equation}\label{eq:Ntraces}
\begin{aligned}
\tfrac{1}{3}\textstyle\sum_{i}\mathcal{N}_{ii}
&=4z(z+Q)+\tfrac{4}{3}\bk^{2}+4|m|^{2},\\
\mathcal{N}_{44}&=-4z(z+Q)+4\bk^{2}+4|m|^{2},
\end{aligned}
\end{equation}
and both need the same two frequency integrals. For ${\rm Re}\,E,{\rm Re}\,F>0$ the poles
$z=iE$ and $z=-Q+iF$ lie in the upper half plane while $z=-iE$ and $z=-Q-iF$ lie in the
lower one, and closing above,
\begin{equation}\label{eq:Ifreq}
\begin{aligned}
I&=\int\!\frac{dz}{2\pi}\frac{1}{(z^{2}+E^{2})[(z+Q)^{2}+F^{2}]}\\
&=\frac{E+F}{2EF\big[(E+F)^{2}+Q^{2}\big]} ,
\end{aligned}
\end{equation}
the apparent pole at $Q=-i(E-F)$ cancelling between the two residues. For the numerator
use $2z(z+Q)=(z^{2}+E^{2})+[(z+Q)^{2}+F^{2}]-(Q^{2}+E^{2}+F^{2})$, which gives
\begin{equation}\label{eq:Kfreq}
K=\int\!\frac{dz}{2\pi}\frac{z(z+Q)}{(z^{2}+E^{2})[(z+Q)^{2}+F^{2}]}
=\frac{E+F}{2\big[(E+F)^{2}+Q^{2}\big]},
\end{equation}
since $\tfrac{1}{4E}+\tfrac{1}{4F}-\tfrac{1}{2}(Q^{2}+E^{2}+F^{2})I$ collapses by
$(E+F)^{2}+Q^{2}-Q^{2}-E^{2}-F^{2}=2EF$. At $F=E^{*}$ one has $E+F=2\varepsilon_{k}$ and
$EF=D_{k}$, so $K=\varepsilon_{k}/(x+4\varepsilon_{k}^{2})$ and
$I=K/D_{k}$. Adding the two orderings, which give the same frequency integral,
\begin{equation}\label{eq:ALradial}
A(x)=\int\!\frac{d^{3}k}{(2\pi)^{3}}\frac{8\varepsilon_{k}T_{+-}}{x+4\varepsilon_{k}^{2}},
\qquad
L(x)=-\int\!\frac{d^{3}k}{(2\pi)^{3}}\frac{8\varepsilon_{k}T_{0}}{x+4\varepsilon_{k}^{2}},
\end{equation}
with $T_{+-}$ of \eqref{eq:Tpm} and $T_{0}=1-(\bk^{2}+|m|^{2})/D_{k}$.

Passing to $s=4\varepsilon_{k}^{2}$ with \eqref{eq:kofeps} and \eqref{eq:lambda}, the
measure is $d^{3}k/(2\pi)^{3}=(2\pi^{2})^{-1}k\,(D_{k}/\varepsilon_{k})\,d\varepsilon_{k}$
and $ds=8\varepsilon_{k}d\varepsilon_{k}$, so that both form factors acquire the same
kernel $1/(s+x)$ with
\begin{equation}\label{eq:rhodef}
\rho_{1,0}(s)=\frac{k\,D_{k}T_{+-,0}}{2\pi^{2}\varepsilon_{k}}
\bigg|_{\varepsilon_{k}=\sqrt{s}/2}.
\end{equation}
The two residues factorize. The first is \eqref{eq:DT1}; the second collapses even more
simply, every term in $\varepsilon_{k}^{2}$ cancelling,
\begin{equation}\label{eq:DT0}
\begin{aligned}
D_{k}T_{0}&=D_{k}-\bk^{2}-|m|^{2}
=-2a^{2}v^{2}\,\frac{\varepsilon_{k}^{2}-N^{2}}{\varepsilon_{k}^{2}}\\
&=-\frac{2a^{2}v^{2}(s-4N^{2})}{s},
\end{aligned}
\end{equation}
which with $k=\sqrt{\lambda}/2\sqrt{s}$ gives \eqref{eq:rhos}. The threshold is at
$s=4N^{2}$ for both, independently of $av$, as Appendix \ref{app:cuts} requires; the
second zero of $\lambda$ sits at $s=-4a^{2}v^{2}$ and opens no cut on this sheet.

The representations need subtracting. Since $\rho_{1}(s)=s/6\pi^{2}+a^{2}v^{2}/\pi^{2}
+O(s^{-1})$ and $\rho_{0}(s)=-a^{2}v^{2}/\pi^{2}+O(s^{-1})$, one subtraction suffices for
$L$ and two are needed for $A$; using
$(s+x)^{-1}-s^{-1}+xs^{-2}=x^{2}s^{-2}(s+x)^{-1}$,
\begin{equation}\label{eq:disprel}
\widehat A(x)=x^{2}\!\!\int_{4N^{2}}^{\infty}\!\!\frac{\rho_{1}(s)\,ds}{s^{2}(s+x)},
\qquad
\widehat L(x)=-x^{2}\!\!\int_{4N^{2}}^{\infty}\!\!\frac{\rho_{0}(s)\,ds}{s^{2}(s+x)},
\end{equation}
with $\widehat A=A-A(0)-xA'(0)$ and likewise for $L$, the subtracted terms being local
contact polynomials. Differentiating twice removes them,
\begin{equation}\label{eq:ddspec}
A''(x)=2\!\!\int_{4N^{2}}^{\infty}\!\!\frac{\rho_{1}(s)\,ds}{(s+x)^{3}},
\qquad
L''(x)=-2\!\!\int_{4N^{2}}^{\infty}\!\!\frac{\rho_{0}(s)\,ds}{(s+x)^{3}} .
\end{equation}
As a check independent of the whole radial route, the same two quantities follow from
\eqref{eq:AL} evaluated with Feynman parameters in dimensional regularization. Writing
$\mathcal{A}(u)=(1-u)m^{2}+u(m^{*})^{2}=N^{2}-a^{2}v^{2}+2iNav(1-2u)$, whose real part is
positive, and using
$I_{2}(\Delta)=-\tfrac{d}{d-2}\Delta I_{0}(\Delta)$ to reduce the tensor integral, one
obtains
\begin{equation}\label{eq:ddparam}
A''(x)=\frac{1}{2\pi^{2}}\!\int_{0}^{1}\!\!du\,
\frac{(u\bar u)^{2}\big(3\mathcal{A}+|m|^{2}+2u\bar u\,x\big)}
{(\mathcal{A}+u\bar u\,x)^{2}},
\end{equation}
with $\bar u=1-u$, and $L''$ the same with the numerator replaced by
$(|m|^{2}-\mathcal{A})(u\bar u)^{2}$; the imaginary parts cancel between $u$ and $1-u$.
We have verified \eqref{eq:ddspec} against \eqref{eq:ddparam} numerically, to a relative
accuracy of $3\times10^{-7}$ over a grid in $N$, $av$ and $x$, which ties the covariant
and the radial computations together and is the sharpest test of \eqref{eq:rhos} we know.

\section{The four-point reflection test}
\label{app:four}

We record the computation behind \eqref{eq:four-negative}. Coordinates are ordered
$(x_{4},x_{1},x_{2},x_{3})$, the parameters are $N=1$, $av=3/5$, $T=\mu=0$, and
${\rm Re}\,m^{2}=16/25>0$, so that the Bessel branches are the continuations from
positive real mass. With $\Psi=(\Psi_{m},\Psi_{m^{*}})^{T}$ and
$V_{a}=i\gamma^{E}_{a}\tau_{x}$, $\tau_{x}$ the flavor exchange, the coordinate
propagator is
\begin{equation}\label{eq:coordS}
\begin{aligned}
S(x)&={\rm diag}\big(S_{m}(x),S_{m^{*}}(x)\big),\\
S_{m}(x)&=(m-\gamma^{E}\!\cdot\!\partial)D_{m}(x),\qquad
D_{m}(x)=\frac{m K_{1}(mr)}{4\pi^{2}r},
\end{aligned}
\end{equation}
$r=|x|>0$. Writing $z=mr$ and using $(K_{1}/z)'=-K_{2}/z$ and
$(-K_{2}/z)'=-K_{2}/z^{2}+K_{3}/z$, the radial derivatives of $D$ follow in closed form,
$D=(m^{2}/4\pi^{2})f$, $D'=(m^{3}/4\pi^{2})f'$ and so on, with
$f=K_{1}/z$ and
$f'''=2K_{2}/z^{3}-K_{2}'/z^{2}+K_{3}'/z-K_{3}/z^{2}$,
$K_{2}'=-(K_{1}+K_{3})/2$, $K_{3}'=-(K_{2}+K_{4})/2$. On the time axis
\begin{equation}\label{eq:axisders}
\begin{gathered}
\partial_{4}D=D'\varsigma,\quad
\partial_{4}^{2}D=D'',\quad
\partial_{i}\partial_{j}D=\frac{D'}{r}\delta_{ij},\\
\partial_{4}^{3}D=D'''\varsigma,\qquad
\partial_{4}\partial_{i}\partial_{j}D
=\Big(\frac{D''}{r}-\frac{D'}{r^{2}}\Big)\delta_{ij}\varsigma,
\end{gathered}
\end{equation}
with $\varsigma={\rm sgn}(x_{4})$, every other component vanishing, and no propagator is
differentiated more than twice.

\begin{table}[b]
\caption{The three-term reduction \eqref{eq:threeterm}: every propagator that occurs,
with any admissible derivative content, in terms of $D$ and its radial derivatives at the
separation $r$, with $\varsigma={\rm sgn}(x_{4})$ and
$\Delta\equiv D''/r-D'/r^{2}$.}
\label{tab:dict}
\begin{ruledtabular}
\begin{tabular}{lccc}
 & $A$ & $B$ & $C$ \\
\colrule
$S$ & $mD$ & $-D'\varsigma$ & $0$ \\
$\partial_{4}S$ & $mD'\varsigma$ & $-D''$ & $0$ \\
$\partial_{1}S$ & $0$ & $0$ & $-D'/r$ \\
$\partial_{4}\partial_{4}S$ & $mD''$ & $-D'''\varsigma$ & $0$ \\
$\partial_{1}\partial_{1}S$ & $mD'/r$ & $-\Delta\varsigma$ & $0$ \\
$\partial_{4}\partial_{1}S$ & $0$ & $0$ & $-\Delta\varsigma$ \\
\end{tabular}
\end{ruledtabular}
\end{table}

This is more than bookkeeping: it collapses the Dirac content of the whole calculation
into a finite table. Since $S=(m-\gamma\!\cdot\!\partial)D$ and, by \eqref{eq:axisders},
only $\partial_{4}$ and one spatial direction ever survive on the axis, \emph{every}
propagator occurring in the test --- undifferentiated, once or twice differentiated ---
has the three-term form
\begin{equation}\label{eq:threeterm}
P=A\,\mathbf{1}+B\,\gamma_{4}+C\,\gamma_{1},
\end{equation}
with the scalars of Table \ref{tab:dict}. The Dirac algebra is therefore exhausted by the
four traces ${\rm Tr}[\gamma_{b_{1}}P_{1}\gamma_{b_{2}}P_{2}]$ and the sixteen traces
${\rm Tr}[\gamma_{b_{1}}P_{1}\cdots\gamma_{b_{4}}P_{4}]$, $b_{i}\in\{4,1\}$, computed once
as polynomials in $A_{1}\dots C_{4}$; the two-point ones are
\begin{equation}\label{eq:tr2}
\begin{aligned}
{\rm Tr}[\gamma_{4}P_{1}\gamma_{4}P_{2}]&=4(A_{1}A_{2}+B_{1}B_{2}-C_{1}C_{2}),\\
{\rm Tr}[\gamma_{1}P_{1}\gamma_{1}P_{2}]&=4(A_{1}A_{2}-B_{1}B_{2}+C_{1}C_{2}),\\
{\rm Tr}[\gamma_{4}P_{1}\gamma_{1}P_{2}]&=4(B_{1}C_{2}+B_{2}C_{1}),
\end{aligned}
\end{equation}
and the four-point ones carry twenty or twenty-one monomials each. Assembled with the
combinatorics below they reproduce \eqref{eq:four-negative} exactly, which is how we have
checked the reduction.

The flavor trace is where the structure of the model reappears. The vertices carry
$\tau_{x}$ and the propagator is diagonal, so for any cycle
\begin{equation}\label{eq:alternate}
\begin{aligned}
&{\rm Tr}_{\rm fl}\big[\tau_{x}S^{(1)}\tau_{x}S^{(2)}\cdots\big]\\
&\qquad=S^{(1)}_{m^{*}}S^{(2)}_{m}S^{(3)}_{m^{*}}\cdots+(m\leftrightarrow m^{*}),
\end{aligned}
\end{equation}
since $\tau_{x}\,{\rm diag}(S_{m},S_{m^{*}})$ is off-diagonal and a product of such
factors is diagonal only for an even number of them. Three things follow at once. Cycles
of odd length vanish. The masses \emph{alternate} around every surviving loop, which is
the balanced-cut condition of Appendix \ref{app:cuts} read off the trace rather than off
the Landau equations. And since the separations are real, $D_{m^{*}}=\overline{D_{m}}$,
so the two terms of \eqref{eq:alternate} are complex conjugates and the result is twice a
real part: the reality of $\omega[(\Theta F)F]$ is structural and not a numerical
accident.

For separated points Wick contraction gives
\begin{equation}\label{eq:wick4}
\begin{aligned}
C_{2}(i,j)&=-{\rm Tr}\big(V_{i}S_{ij}V_{j}S_{ji}\big),\\
C_{4}&=\sum_{\rm pairings}C_{2}C_{2}
-\!\!\sum_{(i,j,k)\in{\rm Perm}(2,3,4)}\!\!L(1,i,j,k),\\
L(1,i,j,k)&={\rm Tr}\big(V_{1}S_{1i}V_{i}S_{ij}V_{j}S_{jk}V_{k}S_{k1}\big),
\end{aligned}
\end{equation}
the trace running over Dirac and flavor indices, with three disconnected pairings and six
oriented four-cycles; cycles of odd length vanish because each vertex carries $\tau_{x}$
while $S$ is flavor diagonal. Applying $\mathcal{O}_{ab}=\partial_{a}j_{b}-\partial_{b}
j_{a}$ at each insertion, the derivative at a point acts on both propagators attached to
it, and the product rule is summed accordingly.

Time reflection acts with $R={\rm diag}(-1,1,1,1)$ on each index, so that
$\Theta\mathcal{O}_{4i}(x)=-\mathcal{O}_{4i}(\theta x)$ and
$\Theta\mathcal{O}_{ij}(x)=+\mathcal{O}_{ij}(\theta x)$; in the electric test the two
reflected insertions supply a combined $+1$. Evaluating at $(-Y,-X,X,Y)$ with
$X=(1,\bm0)$ and $Y=(2,\bm0)$, the disconnected pairing internal to each half is
cancelled by the centering in \eqref{eq:four-witness}, and the two pairings that cross
the reflection together with the connected cycles give the numbers quoted in
\eqref{eq:four-negative}. Smearing against non-negative normalized bumps supported in
disjoint neighbourhoods of $X$ and $Y$, at positive times, preserves the sign, the
reflected and unreflected supports remaining separated so that no contact term enters.
The computation has been carried out twice, with independent implementations of the
Dirac--flavor traces and of the derivatives.

\end{document}